\documentclass{aa}  
\usepackage{natbib}
\usepackage{graphicx}
\usepackage{txfonts}
\usepackage{layouts}
\usepackage[colorlinks,citecolor=blue,linkcolor=blue,urlcolor=blue]{hyperref}
\usepackage{xcolor}
\usepackage{ulem}

\definecolor{chmagenta}{rgb}{0.54, 0.17, 0.88}

\begin{document}

   \title{The $\chi_\mathrm{eff}-z$ correlation of field binary black hole mergers and how 3G gravitational-wave detectors can constrain it}
   \titlerunning{The $\chi_\mathrm{eff}-z$ correlation of field binary black hole mergers and how 3G gravitational-wave detectors can constrain it}

   \author{Simone\,S.\,Bavera\inst{1,2}
          \and
          Maya\,Fishbach\inst{3}
          \and
          Michael\,Zevin\inst{4,5}
          \and
          Emmanouil\,Zapartas\inst{1,6}
          \and
          Tassos\,Fragos\inst{1,2}
          }
    \authorrunning{Bavera et al.}

   \institute{Département d’Astronomie, Université de Genève, Chemin Pegasi 51, CH-1290 Versoix, Switzerland\\
              \email{Simone.Bavera@unige.ch}
         \and 
         Gravitational Wave Science Center (GWSC), Université de Genève, CH1211 Geneva, Switzerland
         \and
             Center for Interdisciplinary Exploration and Research in Astrophysics (CIERA) and Department of Physics and Astronomy, Northwestern University, 1800 Sherman Ave, Evanston, IL 60201, USA
        \and
            Kavli Institute for Cosmological Physics, The University of Chicago, 5640 South Ellis Avenue, Chicago, Illinois 60637, USA
        \and
        Enrico Fermi Institute, The University of Chicago, 933 East 56th Street, Chicago, Illinois 60637, USA
        \and
        IAASARS, National Observatory of Athens, Vas. Pavlou and I. Metaxa, Penteli, 15236, Greece
             }

   \date{Accepted on June 20, 2022 -- ET-0058A-22}

\abstract{
Understanding the origin of merging binary black holes is currently one of the most pressing quests in astrophysics. 
We show that if isolated binary evolution dominates the formation mechanism of merging binary black holes, one should expect a correlation between the effective spin parameter, $\chi_\mathrm{eff}$, and the redshift of the merger, $z$, of binary black holes. 
This correlation comes from tidal spin-up systems preferentially forming and merging at higher redshifts due to the combination of weaker orbital expansion from low metallicity stars given their reduced wind mass loss rate, delayed expansion and have smaller maximal radii during the supergiant phase compared to stars at higher metallicity. As a result, these tightly bound systems merge with short inspiral times.
Given our fiducial model of isolated binary evolution, we show that the origin of a $\chi_\mathrm{eff}-z$ correlation in the detectable LIGO--Virgo binary black hole population is different from the intrinsic population, which will become accessible only in the future by third-generation gravitational-wave detectors such as Einstein Telescope and Cosmic Explorer. 
Given the limited horizon of current gravitational-wave detectors, $z\lesssim 1$, highly rotating black hole mergers in the LIGO--Virgo observed $\chi_\mathrm{eff}-z$ correlation are dominated by those formed through chemically homogeneous evolution. This is in contrast to the subpopulation of highly rotating black holes in the intrinsic population, which is dominated by tidal spin up following a common evolve event. The different subchannel mixture in the intrinsic and detected population is a direct consequence of detector selection effects, which allows for the typically more massive black holes formed through chemically homogeneous evolution to be observable at larger redshifts and dominate the LIGO--Virgo sample of spinning binary black holes from isolated evolution at $z>0.4$.
Finally, we compare our model predictions with population predictions based on the current catalog of binary black hole mergers and find that current data favor a positive correlation of $\chi_\mathrm{eff}-z$ as predicted by our model of isolated binary evolution.
}


\maketitle

%-------------------------------------------------------------------

\section{Introduction}\label{sec:introduction}

%Figures: two columns width \printinunitsof{cm}\prntlen{\textwidth} = 7.2430629921 inches \\
The detection of gravitational waves (GWs) from coalescing binary black holes (BBHs) by the LIGO--Virgo--KAGRA (LVK) collaboration has opened a new window for the study of stellar and binary astrophysics \citep{2015CQGra..32g4001L,2015CQGra..32b4001A,2021PTEP.2021eA101A}. To date, the LVK collaboration has reported 69 BBH events with a false alarm rate (FAR) smaller than $1\,\mathrm{yr}^{-1}$ \citep{2019PhRvX...9c1040A,2021PhRvX..11b1053A,2021arXiv210801045T,2021arXiv211103606T}. However, after more than half a decade since the first detection of GWs, the origin of merging BBHs remains an open question. This is not due to a lack of theoretical predictions but rather because of the degeneracy between different formation channel model predictions and unconstrained astrophysical processes of these models \citep[see, e.g.,][]{2021arXiv210714239M,2021ApJ...910..152Z}.

Improved sensitivity of the LVK detectors and planned third-generation (3G) GW detectors such as the Einstein Telescope \citep{2010CQGra..27s4002P} and the Cosmic Explorer \citep{2019BAAS...51g..35R} will increase BBH detection rates by orders of magnitude. A larger sample size allows for detailed investigations of correlations between BBH observable properties \citep[e.g.,][]{2020JCAP...03..050M,2021arXiv211113991T}, which might enable different astrophysical formation channels to be distinguished. For example, multiple studies have looked for potential correlations between masses and redshifts~\citep{2021ApJ...912...98F,2021arXiv211103634T}, $M_\mathrm{chirp}-\chi_\mathrm{eff}$~\citep{2020ApJ...894..129S,2021arXiv211103634T,2022arXiv220113098F}, $\chi_\mathrm{eff}-q$ \citep{2021ApJ...922L...5C,2021arXiv211103634T}, and $\chi_\mathrm{eff}-z$ \citep{2022arXiv220401578B}. The redshift at which the BBH systems merge, $z$, is a proxy for the distance to the source, $M_\mathrm{chirp}=(m_1m_2)^{3/5}/(m_1+m_2)^{1/5}$ is the chirp mass where $m_1$ and $m_2$ are the BH component masses, $q=m_2/m_1$ is the mass ratio defined with $m_2<m_1$, and $\chi_\mathrm{eff} = (m_1 {\bf{a}}_1 + m_2 {\bf{a}}_2)/(m_1+m_2) \cdot {\bf{\hat{L}}}$ is the effective spin parameter where $\bf{a}_1$ and $\bf{a}_2$ are the component BH dimensionless spin vectors and $\bf{\hat{L}}$ the orbital angular momentum unit vector. 

Here, we demonstrate that field-formed BBHs naturally predict a $\chi_\mathrm{eff}-z$ correlation. Under the assumptions of efficient angular momentum transport inside stars, supported by asteroseismology observational constraints \citep{2014MNRAS.444..102K,2014A&A...564A..27D,2018A&A...616A..24G} and current GW observations \citep{2020A&A...636A.104B,2021ApJ...910..152Z}, and Eddington limited mass accretion efficiency onto BHs, the origin of BH spin in field BBHs arises from tidal interactions during the late BH--Wolf-Rayet (BH-WR) \citep{2018A&A...616A..28Q,2020A&A...635A..97B,2022MNRAS.511.3951F} or WR-WR \citep{2017ApJ...842..111H,2021ApJ...921L...2O} evolutionary phases or, alternatively, through chemically homogeneous evolution induced by rotational mixing caused from tidal spin-up during the early evolutionary stage of close binaries \citep{2016MNRAS.458.2634M,2016A&A...588A..50M}. From first principles, such correlation should exist since the strength of tidal interaction steeply depends on the orbital separation \citep{1977A&A....57..383Z,1981A&A....99..126H}, and the distribution of orbital separation pre core collapse evolves with redshift. The redshift evolution of the orbital separation distribution originates from the metallicity-dependent stellar winds \citep{2000A&A...360..227N,2001A&A...369..574V} whose intensity increases as a function of metallicity. Stronger wind-mass loss (during the BH-WR binary evolution phase) widens the binary more efficiently, inhibiting or even canceling the effects of tides. Because the mean metallicity of the Universe decreases as a function of redshift \citep{2014ARA&A..52..415M,2017ApJ...840...39M}, we empirically expect an increasing fraction of BBH mergers with highly spinning BH components from tidal spin up as a function of redshift. Additionally, low metallicity stars are more compact at zero-age-main-sequence (ZAMS), expand later in their evolution and have smaller maximal radii during the supergiant phase compare to stars at higher metallicity.

In this paper, we discuss the evolving $\chi_\mathrm{eff}$ distribution as a function of redshift for field BBHs from the common envelope (CE), stable mass transfer (SMT), and the chemically homogeneous evolution (CHE) channels. 
%was implicitly shown in \citet{2022A&A...657L...8B}.
%There, it was investigated how the formation of highly spinning merging BBHs could lead to the emission of luminous long-duration gamma-ray bursts (LGRBs). If indeed such channel would dominate LGRB emission, then LGRBs could be used to probe the formation of highly rotating merging BBHs up to high redshifts ($z\simeq 9$), outside current GW detectors horizons.  Nevertheless, we believe that the result of a $\chi_\mathrm{eff}-z$ correlation of field BBHs is of such importance that we want to highlight it in this manuscript explicitly and discuss its direct implication to the current and future field of GW astrophysics.
The paper is structured as follows. First, we introduce our fiducial model and describe how we quantify the $\chi_\mathrm{eff}-z$ correlation in Section~\ref{sec:methods}. In Section~\ref{sec:results}, we present the redshift evolution of the $\chi_\mathrm{eff}$ distribution for the intrinsic and detectable BBH population as predicted by our model of isolated binary evolution. We then compare our model predictions against the LIGO--Virgo catalog of BBHs. In Section~\ref{sec:discussion}, we discuss how potential uncertainties in our model might affect the $\chi_\mathrm{eff}$ distribution of field BBHs and how a possible change in the mixing fraction of the different channels predicted from isolated binary evolution might impact our results. All findings are summarised in Section~\ref{sec:conclusions}.

\section{Methods}\label{sec:methods}

\subsection{The binary black-hole population synthesis model}
This study uses the isolated binary evolution model presented in \citet{2022A&A...657L...8B}, calculated using the \texttt{POSYDON} framework \citep{2022arXiv220205892F}, which accounts for BBH formation through the CE, SMT, and CHE channels. It was shown that this model (i) leads to BBH observable properties consistent with the events of the second LIGO--Virgo GW transient catalog (GWTC-2)  \citep{2021ApJ...910..152Z}, (ii) have BBH merger rate estimates compatible with observational constraints of GWTC-2 and, now GWTC-3, \citep{2021A&A...647A.153B, 2020MNRAS.499.5941D}, (iii) the subpopulation of highly spinning BBHs might explain the observed population of luminous LGRBs across the cosmic history of the Universe \citep{2022A&A...657L...8B}, and (iv) does not violate current upper limit estimates of the stochastic GW background \citep{2021arXiv210905836B}. 

In contrast to most rapid population synthesis studies, our simulations accurately model the late tidal spin-up phase of the second-born BH and CHE due to rotational-induced mixing of tidally spun-up ZAMS binaries. The former is done by following the evolution of the binaries from ZAMS up to the formation of the BH-WR systems after the second mass transfer phase with the rapid population synthesis code \texttt{COSMIC} \citep{2020ApJ...898...71B} and then uses detailed \texttt{MESA} \citep{2011ApJS..192....3P,2013ApJS..208....4P, 2015ApJS..220...15P,2018ApJS..234...34P,2019ApJS..243...10P} BH-WR simulations \citep{2021A&A...647A.153B} to accurately model the final tidal spin-up phase of the BH-WR system up to central carbon exhaustion of the WR star, as done in \citet{2022A&A...657L...8B}.
The detailed BH-WR simulations self-consistently model the angular momentum evolution of the WR star, which is determined by the interplay of tides, WR stellar wind mass loss, and the evolution of the WR stellar structure.

Massive stars in short orbital periods ($p<2\,\mathrm{days}$) at ZAMS with nearly equal masses tidally spin up to be highly rotating, which induces rotational mixing and eventually leads to CHE. 
Because \texttt{COSMIC} cannot model the parameter space leading to CHE as the code cannot accurately follow the back-reaction on the stellar structure and evolution from rotational-induced mixing,
CHE is done by matching ZAMS binary conditions to detailed \texttt{MESA} simulations targeting CHE according to \citet{2020MNRAS.499.5941D}, as implemented in \citet{2022A&A...657L...8B}.

Given the availability of the stellar profile at carbon exhaustion from the \texttt{MESA} simulations, in both cases, the core collapse considers disk formation during the collapse of highly spinning stars. Additionally, we account for mass loss through neutrinos, pulsational pair-instability and pair-instability supernovae (PPISNe \& PISNe) \citep{2019ApJ...882...36M}, and orbital changes resulting from anisotropic mass loss and isotropic neutrinos mass loss \citep{1996ApJ...471..352K}, as explained in Appendix D of \citet{2021A&A...647A.153B}. Because we implement the \citet{2012ApJ...749...91F} delayed collapse mechanism which assigns zero velocity kicks to collapsing stars with carbon-oxygen cores with masses above $11\,M_\odot$, in practice, we find a statistically small number of systems with $\chi_\mathrm{eff}<0$. Alternatively, non-negligible kicks would lead to a more considerable fraction of negative $\chi_\mathrm{eff}$ \citep[see, e.g.,][]{2016ApJ...832L...2R,2018PhRvD..98h4036G,2021ApJ...920..157C,2022ApJ...926L..32S}.
For a detailed explanation of the main features and the physical assumptions made in this model, we refer the reader to the extensive discussions in \citet{2022A&A...657L...8B}. 

\subsection{Detection rates}
Merger rates are computed by convolving the redshift and metallicity dependent star formation rate as predicted by the Illustris-TNG simulation \citep{2015A&C....13...12N} with the synthetic catalog of merging BBHs obtained by evolving initial ZAMS conditions at different discrete metallicities with $\texttt{POSYDON}$. Following the notation of \citet{2020A&A...635A..97B,2021A&A...647A.153B,2022A&A...657L...8B}, the BBH detection rate of a GW detector network can be expressed as a Monte Carlo sum over the synthetic population of merging BBHs, i.e., $R_\mathrm{det} = \sum_{i,j,k} w_{i,j,k}(p_\mathrm{det}) \, \mathrm{yr}^{-1}$ where $w_{i,j,k}$ is the weighted contribution of a binary $k$ forming at redshift $z_{\mathrm{f},i}$ and merging at redshift $z_{\mathrm{m},k} \equiv z_k$. Here the dummy index $j$ indicates the discrete sum over the 30 simulated log-binned metallicity intervals $\Delta Z_j$. The synthetic BBH population is distributed across the cosmic history of the Universe in the center of time bins of size $\Delta t_i = 100\,\mathrm{Myr}$ with center the formation redshift $z_{\mathrm{f},i}$. We chose the time bin size to be small enough to ensure the convergence of our results (see Appedinx~D of \citealt{2022A&A...657L...8B} for the details of the calculation). 

To compute the BBH detection rate of LIGO--Virgo, we account for the detectors' selection effects, $p_\mathrm{det}$, given the source redshift, BH masses, and spins. Here, we assume a GW detector network configuration composed by LIGO Hanford, LIGO Livingston, and Virgo at O3 \texttt{mid-high/late-low} sensitivity \citep{2018LRR....21....3A} with a network signal-to-noise ratio (S/N) threshold of 12 as implemented in \citet{2021A&A...647A.153B}. 

We also consider the sensitivity of the future 3G ground-based GW detector Einstein Telescope. To approximate the BBH detection rate of the Einstein Telescope, we account for detector selection effects given the source redshift and BH masses assuming a theorized noise-sensitive curve \texttt{ET-D} \citep{2011CQGra..28i4013H} as implemented by \citet{2018MNRAS.477.4685B} in \texttt{COMPAS} \citep{2021arXiv210910352T}. Here, we assume a conservative S/N threshold of 12 for the Einstein Telescope, similar to what \citet{2011CQGra..28i4013H} assumed. In practice, we find that this assumption sets the horizon of a BBH with $m_1=m_2=15\,M_\odot$ at $z=10$, namely $p_\mathrm{det}^\mathrm{ET}(z=10,m_1=m_2=15\,M_\odot)\simeq 0$.

Finally, we will distinguish the intrinsic detection rate, i.e., what a GW detector with infinite sensitivity would observe on Earth, using the notation $\tilde{w}_{i,j,k}=w_{i,j,k}(p_\mathrm{det}=1)$ as first introduced in \citet{2022A&A...657L...8B}.

\subsection{Relative channel contribution}

Once detection rates are defined, we can compute the relative contribution to the intrinsic detection rate from each one of the isolated binary evolution channels (CE, SMT, and CHE) at a given redshift as
\begin{equation}
    f_{\rm channel} (z) = \sum_{i,j,k}  \frac{\tilde{w}_{i,j,k}(k\,|\,k\in\mathrm{channel}) }{\tilde{w}_{i,j,k}} \Bigg|_{z_k \in \Delta z}
    \label{eq:f_channel}
\end{equation}
where $z\in[0,10]$ is discredited in bins, $\Delta z$, taken to have a constant cosmic time width of $\Delta t = 200\,\mathrm{Myr}$. Similarly, for the detectable population, we define $f^\mathrm{det}_{\rm channel} (z)$ where we use $w_{i,j,k}$ instead of $\tilde{w}_{i,j,k}$ in $z\in[0,1]$ for LIGO--Virgo and $z\in[0,10]$ for the Einstein Telescope.

\subsection{Quantifying the $\chi_\mathrm{eff}-z$ correlation}

To quantify the redshift evolution of the $\chi_\mathrm{eff}$ distribution, we define $f_{\chi_\mathrm{eff}>\chi_0}(z)$ to be the fraction of merging BBHs with $\chi_\mathrm{eff}$ above the arbitrary value $\chi_0$ at a given redshift for the modeled intrinsic BBH population. This quantity is calculated as
\begin{equation}
    f_{\chi_\mathrm{eff}>\chi_0} (z) = \sum_{i,j,k}  \frac{\tilde{w}_{i,j,k}(\chi_{\mathrm{eff},k}|\chi_\mathrm{eff}>\chi_0) }{\tilde{w}_{i,j,k}} \Bigg|_{z_k \in \Delta z}
    \label{eq:f_chi}
\end{equation}
 and, similarly, for the detectable populations, we define $f^\mathrm{det}_{\chi_\mathrm{eff}>\chi_0}(z)$ with the same redshift spacing and bounds as in Eq.~\eqref{eq:f_channel}.

\begin{figure*}
\centering
\includegraphics{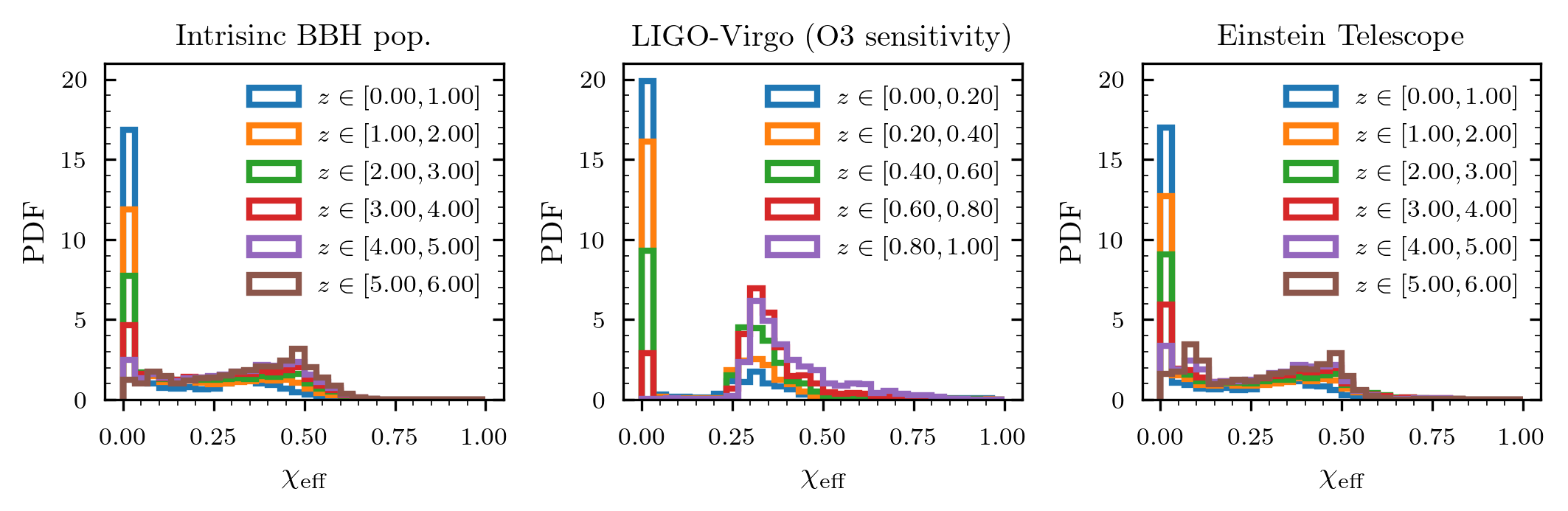}
\caption{Effective spin parameter, $\chi_\mathrm{eff}$, distribution of field BBHs as a function of redshift, z. (\textit{Left}) We show the modeled intrinsic (underlying) population of field merging BBHs. (\textit{Center}) We show the modeled detectable LIGO--Virgo BBH population assuming simulated O3 detector sensitivity selection effects. (\textit{Right}) We show the modeled Einstein Telescope detectable BBH population assuming a forecast detector sensitivity as in \citet{2011CQGra..28i4013H}. In all cases, the fraction of non-spinning BBHs decreases as a function of redshift, shifting the $\chi_\mathrm{eff}$ distribution to larger $\chi_\mathrm{eff}$ values.}
\label{fig:chi_eff_model}
\end{figure*}

\section{Results}\label{sec:results}

We now investigate the $\chi_\mathrm{eff}-z$ correlation of field-formed BBHs and assert its detectability given current and planned GW observatories. We first look at the intrinsic and detectable $\chi_\mathrm{eff}$ distributions as a function of redshift in our fiducial model, which includes potential contribution from the CE, SMT, and CHE channels described in Section~\ref{sec:chi_eff_evol}. We then quantify the intrinsic and detectable $\chi_\mathrm{eff}-z$ correlation by computing the quantities $f_{\chi_\mathrm{eff}>\chi_0}(z)$ and $f^\mathrm{det}_{\chi_\mathrm{eff}>\chi_0}(z)$ and the relative channel contributions $f_\mathrm{channel}(z)$ and $f^\mathrm{det}_\mathrm{channel}(z)$, in Section~\ref{sec:chi_z_correlation}. Finally, we look for evidence of the modeled $f_{\chi_\mathrm{eff}>\chi_0}(z)$  and $f^\mathrm{det}_{\chi_\mathrm{eff}>\chi_0}(z)$ in LIGO--Virgo GWTC-3 data in Section~\ref{sec:GWTC-3_corr}.

\begin{figure*}
\centering
\includegraphics{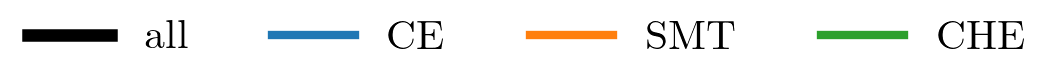}\\
\includegraphics{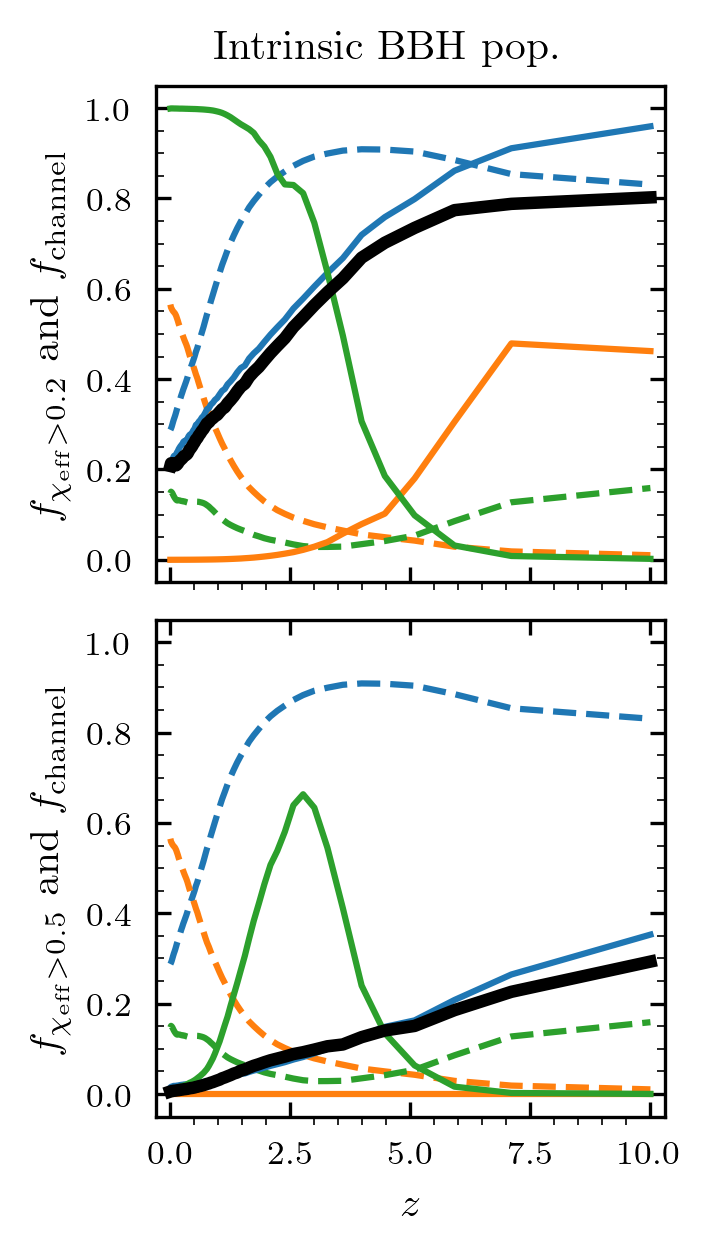} %[width=0.495\linewidth]
\includegraphics{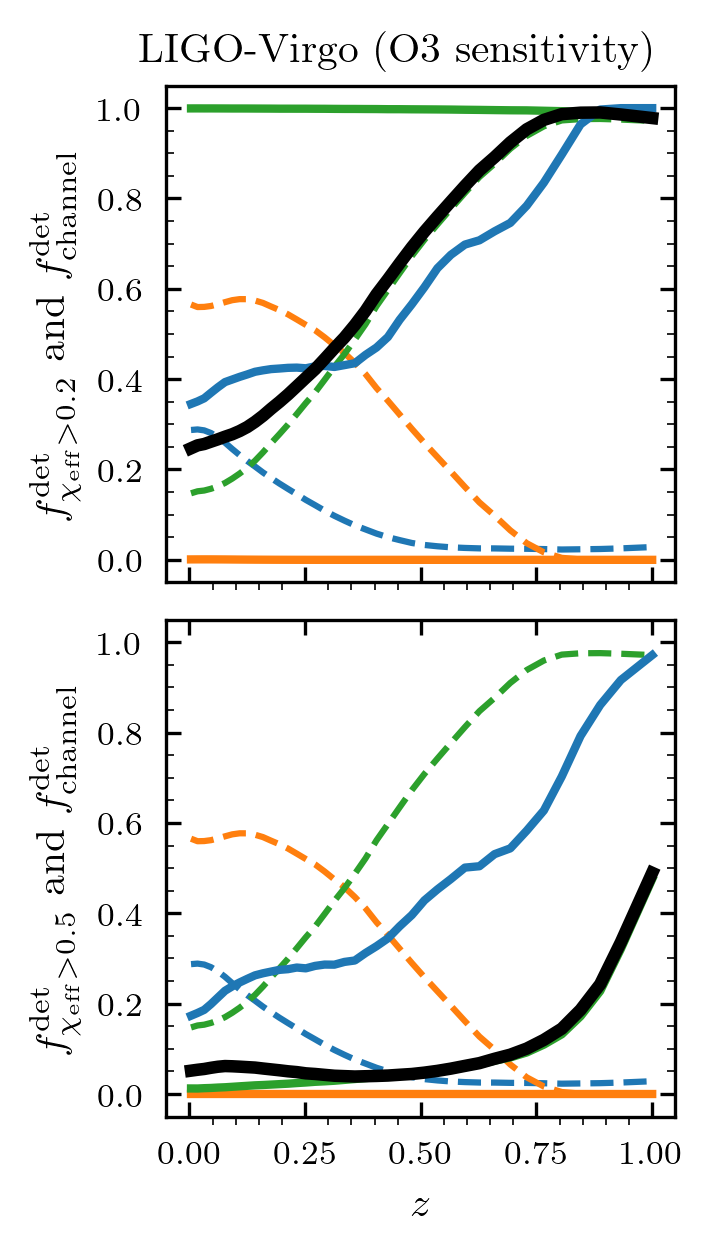}
\includegraphics{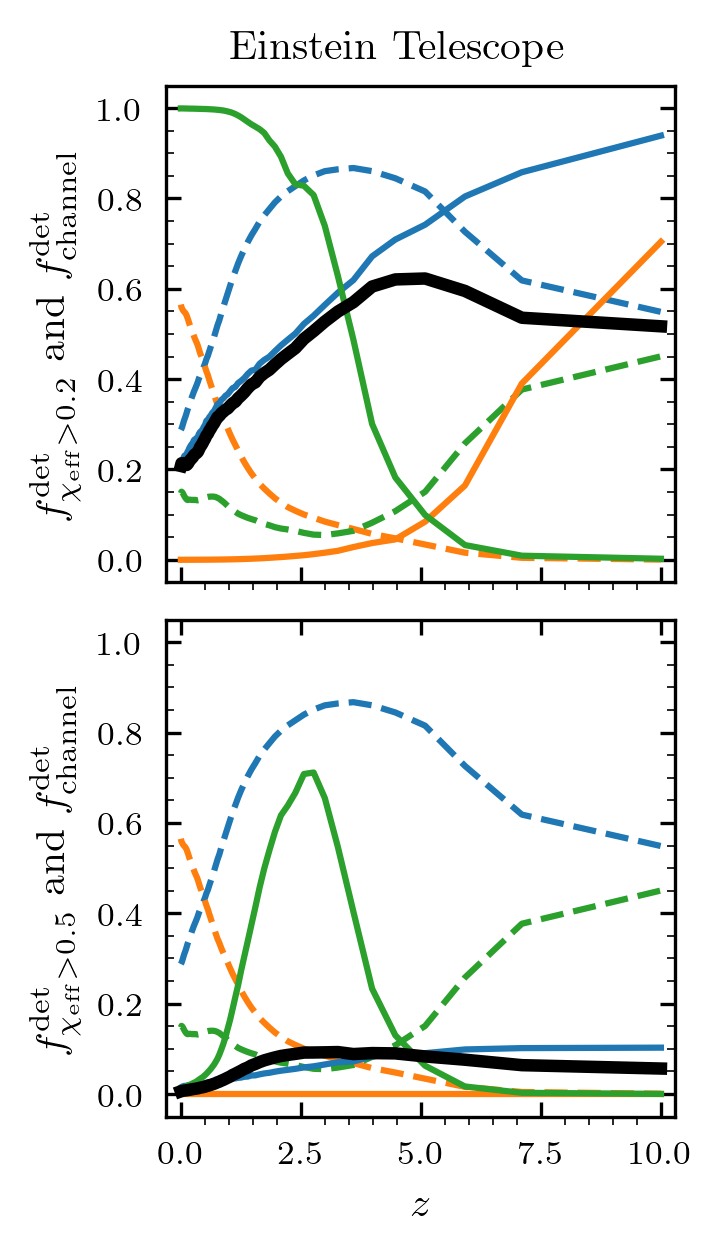}
\caption{Fractions of BBHs with $f_{\chi_\mathrm{eff} > 0.2}$ and $f_{\chi_\mathrm{eff} > 0.5}$ as a function of redshift (solid lines), and the relative contribution of each field BBH channel, $f_\mathrm{channel}$, according to the legend (dashed lines). Black solid lines show the fraction of systems that satisfy a given $\chi_\mathrm{eff}$ criteria for the combined CE, SMT, and CHE channels with redshift-dependent branching fractions accounted for. (\textit{Left}) We show the modeled intrinsic (underlying) population of field merging BBHs. (\textit{Center}) We show the modeled LIGO--Virgo detectable BBH population assuming simulated O3 detector sensitivity selection effects. (\textit{Right}) We show the modeled Einstein Telescope detectable BBH population assuming forecast detector sensitivity as in \citet{2011CQGra..28i4013H}. In most cases, the fraction of highly spinning BBHs increases as a function of redshift.}
\label{fig:model_f_chi}
\end{figure*}

\subsection{The $\chi_\mathrm{eff}$ distribution of field BBHs}\label{sec:chi_eff_evol}

First, we show the $\chi_\mathrm{eff}$ distribution as a function of discrete redshift bins for the intrinsic and detectable BBH populations in Figure~\ref{fig:chi_eff_model}. 
%For the detectable BBH population, we consider the LIGO--Virgo GW detector network assuming O3 sensitivity as well as make a conservative forecast of what the Einstein Telescope might observe in the future, assuming the theorised sensitivity study of \citet{2011CQGra..28i4013H}.
At low redshifts, the intrinsic distribution manifests a peak at $\chi_\mathrm{eff}=0$ plus an almost flat distribution up to $\chi_\mathrm{eff}\simeq 0.5$ which then progressively decays. Similarly, the detectable LIGO--Virgo $\chi_\mathrm{eff}$ distribution also exhibit a similar narrow peak at $\chi_\mathrm{eff}=0$. However, in contrast to the intrinsic distribution, we observe a second broader peak at around $\chi_\mathrm{eff}\simeq 0.35$ with an elongated tail reaching large $\chi_\mathrm{eff}$ depending on redshift. Both distributions evolve with redshift; with the $\chi_\mathrm{eff}$ distribution in the intrinsic population showing a slow evolution with redshift, while in the LIGO--Virgo observable population the distribution evolves significantly over the redshift range between 0 and 1. The median $\bar{\chi}_\mathrm{eff}$ value of the intrinsic distribution grows from $\bar{\chi}_\mathrm{eff}^{z\in[0,1]}\simeq 0.12$ to $\bar{\chi}_\mathrm{eff}^{z\in[5,6]} \simeq 0.33$ while for the LIGO--Virgo detectable population the model predicts that $\bar{\chi}_\mathrm{eff}^{z\in[0,0.2]}\simeq 0.13$ grows to $\bar{\chi}_\mathrm{eff}^{z\in[0.8,1]}\simeq 0.41$.

The origin of the redshift evolution of the $\chi_\mathrm{eff}$ distribution is different between the intrinsic and the detectable LIGO--Virgo BBH populations as they probe different redshift horizons, $z\in[0,\infty]$ and $z\in[0,1]$, respectively. The former encapsulates all merging BBHs at any redshifts and probes the increasing fraction of systems experiencing tidal spin up prior to BBH formation at increasing redshifts (see Section~\ref{sec:introduction}).
In contrast, the detectable LIGO--Virgo population is biased by the BBH mass-dependent selection effects.  More massive BHs can be detected at further distances than lighter BHs. We note that at the highest redshifts detectable by LIGO--Virgo ($z\simeq 1$), only systems with large positive $\chi_\mathrm{eff}$ are detected due to the increased duration of the inspiral and therefore the S/N. However, the impact of $\chi_\mathrm{eff}$ on detectability is minor compared to mass selection effects \citep{2018PhRvD..98h3007N}.
% and therefore do not take spins into account for our detectability predictions \citep{2018PhRvD..98h3007N}. 

In the following section, we show how the different evolutionary channels CE, SMT, and CHE, which have distinct spin distributions, have different detector horizons due to their inherent mass spectrum.
For a discussion about the intrinsic and LIGO--Virgo observable joint distributions of $\chi_\mathrm{eff}$ vs. $M_\mathrm{chirp}$, we refer the reader to Figure~1 of \citet{2022A&A...657L...8B}. 
Finally, because the Einstein Telescope has a much more distant horizon than current generation GW detectors, the planned GW observatory will be able to detect the majority of the underlying BBH population up to large redshifts, see e.g., $z \in [4,5]$ in Figure~\ref{fig:chi_eff_model}. We therefore find that the Einstein Telescope will observe an evolving $\chi_\mathrm{eff}$ distribution similar to the intrinsic one.

\subsection{The $\chi_\mathrm{eff}-z$ correlation of field BBHs}\label{sec:chi_z_correlation}

The $\chi_\mathrm{eff}-z$ correlation of field BBHs in the intrinsic population, $f_{\chi_\mathrm{eff}>\chi_0}(z)$, is shown in the leftmost column of Figure~\ref{fig:model_f_chi} for $\chi_0=0.2$ and $\chi_0=0.5$. In both cases, $f_{\chi_\mathrm{eff}>\chi_0}(z)$ is monotonically increasing and reaches an asymptotic plateau at high redshifts, $z>5$ and $z>8$ for $\chi_0=0.2$ and $\chi_0=0.5$, respectively.
To understand the origin of the $f_{\chi_\mathrm{eff}>\chi_0}(z)$ shape, we need to look at this quantity channel-wise and consider the relative contribution of each channel $f_{\rm channel} (z)$ to the total BBH intrinsic population. 
In Figure~\ref{fig:model_f_chi}, we can see that at low redshifts the intrinsic BBH merging population is composed of a mix of channels, $f_{\rm CE} (z=0) = 30\%, f_{\rm SMT}(z=0) = 55\%$, and $f_{\rm CHE}(z=0) = 15\%$. In contrast, at higher redshifts, the total population of merging BBHs is dominated by the CE channel, with $f_{\rm CE}(z\geq 2) \geq 80\%$. 

Channel-wise, we can see that $f^{\rm CE}_{\chi_\mathrm{eff}>\chi_0}(z)$ increases monotonically due to a larger fraction of systems experiencing tidal spin-up as a function of redshift. On average, at higher redshifts, binary systems are born at lower metallicities and experience reduced stellar wind mass loss. Hence, an increased fraction of binaries can maintain short orbital separations and tidal locking during the BH-WR phase.
A similar argument can be made for the SMT channel. However, because SMT leads on average to wider BH-WR orbital separations than the CE channel \citep{2021A&A...647A.153B}; we find $f^{\rm SMT}_{\chi_\mathrm{eff}(z)>\chi_0}(z)<f^{\rm CE}_{\chi_\mathrm{eff}>
\chi_0}(z)$ for any redshift. Moreover, we find that at low redshifts $f^{\rm SMT}_{\chi_\mathrm{eff}>0.2}(z\simeq0) = 0$, which steadily increases to $f^{\rm SMT}_{\chi_\mathrm{eff}>0.2}(z\simeq 10) \simeq 0.5$. 
On the contrary, $f^{\rm CHE}_{\chi_\mathrm{eff}>0.2}(z)$ manifests a monotonically decreasing behavior. At low redshifts, $f^{\rm CHE}_{\chi_\mathrm{eff}>0.2}(z \simeq 0)=1$, i.e. all BBHs from the CHE channel are fast spinning, while at higher redshifts most systems possess negligible $\chi_\mathrm{eff}$, with $f^{\rm CHE}_{\chi_\mathrm{eff}>0.2}(z = 10) = 0$. 
We also notice that highly rotating CHE systems with $\chi_\mathrm{eff}>0.5$ are not present in the local universe $f^{\rm CHE}_{\chi_\mathrm{eff}>0.5}(z=0)=0$ but their presence peaks at $f^{\rm CHE}_{\chi_\mathrm{eff}>0.5}(z=2.5)=0.6$ before decreasing again to $f^{\rm CHE}_{\chi_\mathrm{eff}>0.5}(z=10)=0$.
The decreasing fraction of highly rotating CHE systems as a function of redshift is a direct consequence of angular momentum loss due PPISNe.
The CHE channel only operates at low metallicities ($Z < 5\cdot 10^{-3}$) but only binaries with metallicities $Z\leq 10^{-4}$ experience mass loss due to PPISNe (in the considered ZAMS primary masses range $\leq 150 \,M_\odot$). During these pulses, the mass ejection from the stellar surface depletes the angular momentum content of these stars and leads to slowly rotating BHs, see Appendix~\ref{sec:AM_loss} for more details. Because at large redshifts ($z>3$) the metallicity-dependent star formation rate leads to an increasing relative fraction of extremely low metallicity binaries, we expect to observe a decreasing $f^\mathrm{CHE}_{\chi_\mathrm{eff}>\chi_0}$ as a function of increasing redshift.
The three channels combined lead to the monotonically increasing behavior of $f_{\chi_\mathrm{eff}>\chi_0}(z)$ we see in Figure~\ref{fig:model_f_chi}, which is mainly dominated by the CE channel. In Appendix~\ref{sec:no_channel}, we show how our fiducial model predictions would change if one of these three channels would be neglected, see Section~\ref{sec:discussion} for a discussion of these alternative scenarios.

The $\chi_\mathrm{eff}-z$ correlation of field BBHs in the detectable populations, $f^\mathrm{det}_{\chi_\mathrm{eff}>\chi_0}(z)$, is shown in the center and right columns of Figure~\ref{fig:model_f_chi} for LIGO--Virgo detectors at O3 sensitivity and the Einstein Telescope, respectively, for $\chi_0=0.2$ and $\chi_0=0.5$. For LIGO--Virgo detectability, $f^\mathrm{det}_{\chi_\mathrm{eff}>\chi_0}(z)$ is a monotonically increasing function growing from $f^\mathrm{det}_{\chi_\mathrm{eff}>0.2}(z=0)=0.25$ to $f^\mathrm{det}_{\chi_\mathrm{eff}>0.2}(z=1)=1$, and $f^\mathrm{det}_{\chi_\mathrm{eff}>0.5}(z=0)=0.05$ to $f^\mathrm{det}_{\chi_\mathrm{eff}>0.5}(z=1)=0.5$. 
On the other hand, the Einstein Telescope $f^\mathrm{det}_{\chi_\mathrm{eff}>\chi_0}(z)$  mimics the intrinsic distribution up to $z \simeq 5$ above which it shows a suppression. The similarity between the Einstein Telescope detectable distribution and the underlying distribution is due to the Einstein Telescope GW horizon being much more distant than that of LIGO--Virgo. The suppression for the detectable Einstein Telescope $f^\mathrm{det}_{\chi_\mathrm{eff}>\chi_0}(z>5)$ is caused by the fact that the detector cannot resolve all distant low mass highly rotating BBHs formed through the CE channel.
To understand the difference between the LIGO--Virgo $f^\mathrm{det}_{\chi_\mathrm{eff}>\chi_0}$ redshift evolution compared to the Einstein Telescope and the intrinsic BBH population, we need to once again consider the relative contribution of each channel. 
Similar to the intrinsic distribution, for LIGO--Virgo at low redshift, we have a mixed contribution of the different channels, $f^\mathrm{det}_{\rm CE}(z=0) = 30\%, f^\mathrm{det}_{\rm SMT}(z=0) = 55\%$, and $f^\mathrm{det}_{\rm CHE} (z=0)= 15\%$. Up to redshift $z=0.4$ the SMT channel dominates over CE and CHE, 
above which the CHE channel dominates the LIGO--Virgo detectable population to the point where $f^\mathrm{det}_{\rm CHE}(z\geq 0.75) \simeq 1$. 
This is notably different than the behavior of the intrinsic BBH population and is a direct consequence of selection effects favouring high BH masses. The different channels have different BH mass distributions, which result in different observational horizons for each channel. Notably, the CHE channel leads to more massive BBHs compared to CE and SMT, and hence this channel can be probed by LIGO--Virgo at larger redshifts compared to BBHs formed from the CE and SMT channels. The described signature leads to a bimodal distribution of $\chi_\mathrm{eff}$ in the LIGO-Virgo detectable BBH population in Figure~\ref{fig:chi_eff_model}. The second peak is mostly composed of BBHs formed through the CHE channel \citep[see][for further discussions]{2022A&A...657L...8B}. This bimodal feature is not present in the $\chi_\mathrm{eff}$ distribution of the intrinsic BBH population for $z\in[0,1]$ (see the right panel of Figure~\ref{fig:chi_eff_model}).
Because intrinsically $f^{\rm CHE}_{\chi_\mathrm{eff} > 0.2} (z\leq 1) \simeq 1$, $f^{\rm SMT}_{\chi_\mathrm{eff}> 0.2} (z\leq 1) \simeq 0$, and $f^{\rm CE}_{\chi_\mathrm{eff} > 0.2}(z)$ is monotonically increasing at low redshifts, we also find a monotonically increasing $f^{\rm det}_{\chi_\mathrm{eff}> 0.2}(z)$ function for LIGO--Virgo sensitivity.
A similar argument can be made for $f^{\rm det}_{\chi_\mathrm{eff}> 0.5}(z)$, where for low redshifts, it holds that $f^{\rm SMT}_{\chi_\mathrm{eff}> 0.5} (z\leq 1) \simeq 0$ while both $f^{\rm CHE}_{\chi_\mathrm{eff}> 0.5} (z\leq 1)$ and $f^{\rm CE}_{\chi_\mathrm{eff}> 0.5} (z\leq 1)$ are monotonically increasing, which results in $f^{\rm det}_{\chi_\mathrm{eff}> 0.5}(z)$ monotonically increasing. Finally, notice that the CHE dominance is not present in the Einstein Telescope detectable population as the 3G detector, given our theorised sensitivity curve, will be able to observe the entire intrinsic BBH population up to redshift $z\simeq 4-5$.

\begin{figure}
\centering
\includegraphics{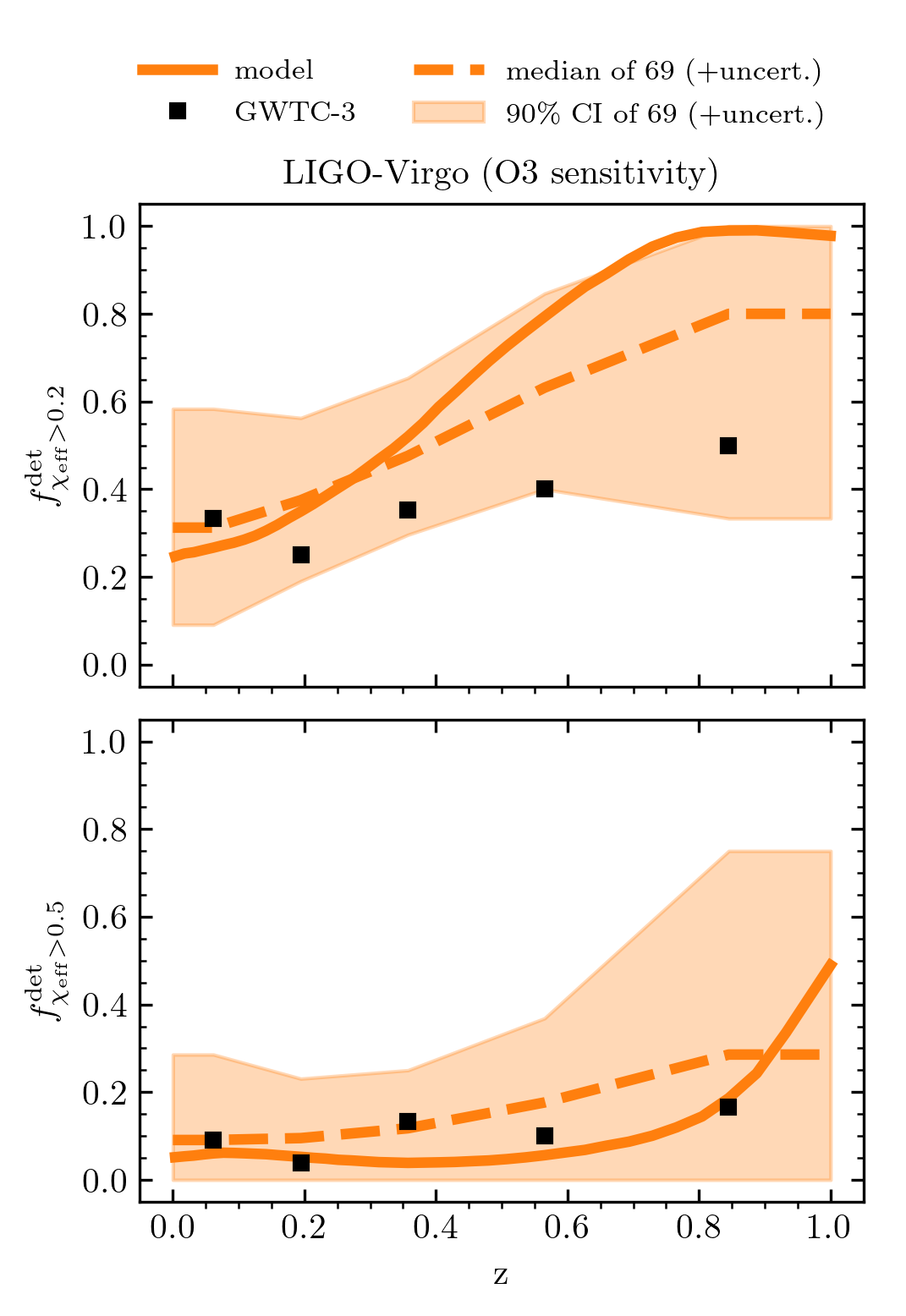}%[width=\linewidth]
\caption{Modeled and observed fractions of BBHs satisfying $f^\mathrm{det}_{\chi_\mathrm{eff} > \chi_0}$ as a function of the redshift. Samples are placed into redshift bins with a bin size of $\Delta t = 1.6\,\mathrm{Gyr}$. The observed fractions $f^\mathrm{det}_{\chi_\mathrm{eff} > 0.2}$ and $f^\mathrm{det}_{\chi_\mathrm{eff} > 0.5}$ are obtained from the median of 10,000 GWTC-3 mock catalog events obtained by sampling the 69 events with $\mathrm{FAR} < 1\,\mathrm{yr}^{-1}$ likelihoods. The modeled prediction for O3 sensitivity is shown with a solid orange line. To compare the model with the data, we generated 10,000 mock catalogs of 69 events, to which we added mock measurements uncertainties. We indicate the median and 90\% CI modeled fractions with orange dashed line and shaded area, respectively. Mock uncertainties are obtained from the zero-centered GWTC-3 event likelihoods.}
\label{fig:model_vs_GWTC3}
\end{figure}

\subsection{Evidence for the $\chi_\mathrm{eff}-z$ correlation in GWTC-3 data}\label{sec:GWTC-3_corr}

Our model provides a falsifiable prediction that both the underlying and detected high-$\chi_\mathrm{eff}$ fractions $f_{\chi_\mathrm{eff}>\chi_0}(z)$ and $f^\mathrm{det}_{\chi_\mathrm{eff}>\chi_0}(z)$ for the O3 LIGO--Virgo detector network should increase as a function of redshift if isolated evolution channels dominate the BBH merger rate at low redshifts. We notice that at low redshifts, $z<1$, the evolution of this fraction for the intrinsic BBH population is mild, but can be amplified by the selection effects of current ground-based detectors. We use BBH events from GWTC-3 to infer both $f_{\chi_\mathrm{eff}>\chi_0}(z)$ and $f^\mathrm{det}_{\chi_\mathrm{eff}>\chi_0}(z)$ and compare them against our model predictions. As in \citet{2021arXiv211103634T}, we only consider GWTC-3 events with a false alarm rate (FAR) smaller than $1\,\mathrm{yr}^{-1}$. In GWTC-3, there are 76 events satisfying this condition from which we exclude the binary neutron stars (NSs) GW170817 and GW190425\_081805, the NS-BH systems GW190426\_152155, GW200105\_162426, GW200115\_042309, and the events GW190814, GW190917\_114630 in which the less massive compact objects have masses that could be either a massive NS or a BH. Our BBH sample therefore includes a total of 69 BBH events.

We first approximate the observed $f^\mathrm{det}_{\chi_\mathrm{eff}>\chi_0}(z)$ for O3, or $f^\mathrm{GWTC-3}_{\chi_\mathrm{eff}>\chi_0}(z)$, in a model agnostic way directly from the observed events. We measure $f^\mathrm{GWTC-3}_{\chi_\mathrm{eff}>\chi_0}(z)$ on a sample of 10,000 mock GWTC-3 catalogs composed of 69 BBH events. A mock catalog of events, $\{x_i\}_{i=1}^{N=69}$, is generated by drawing a 2D sample, $x^k_i=(\chi^k_\mathrm{eff},z^k)_i$, from each event's $x_i$ 2D posterior distribution\footnote{In contrast to the GWTC-3 official analysis, for events in O3b we use posterior samples from the IMRPhenomXPHM analysis as the Mixed and SEOBNRv4PHM analyses do not come with associated prior samples in the 8th November 2021 data release \citep{ligo_scientific_collaboration_and_virgo_2021_5546663}. } $p(\chi_\mathrm{eff},z|x_i)$ weighted by the inverse of the prior 2D probability density $p(\chi_\mathrm{eff},z)$ in order to sample from the likelihood. The events' posterior and prior distributions are publicly released by the LIGO--Virgo collaboration. We approximate the discretely-sampled prior distribution probability density function (PDF) with a 2D kernel density estimator (KDE) trained on the GWTC-3 event samples where the bandwidth of the KDE is set by Scott's rule \citep{2015mdet.book.....S} as implemented in the Gaussian KDE function of the \texttt{SciPy} Python module \citep{2020SciPy-NMeth}. The accuracy of our KDE method to represent the inferred 2D distributions is verified by comparing the histogram of the original samples and samples generated from the KDEs. 

To perform a fair comparison of our model with the observations, we need to account for (i) the statistical variance of drawing a sample of 69 events from our model and (ii) to account for the measurement uncertainty for BBH parameters. This is done by generating 10,000 mock samples of 69 events from our model, to which we add mock uncertainty to each event. We approximate measurement uncertainties following a procedure first shown in \citet{2020A&A...635A..97B} for the $\chi_\mathrm{eff}$ parameter. Here, we extend this procedure to the 2D case. Mock uncertainties are obtained by shifting another set of 10,000 mock GWTC-3 catalogs by each event's median value $\bar{x}_i=(\bar{\chi}_\mathrm{eff},\bar{z})_i$. When the mock uncertainty is added to the model mock samples, we find that this methodology overestimates the measurement uncertainty of events with low redshift of merger. This occurs because this methodology does not assign smaller measurement uncertainties to events with smaller redshifts of merger. Such correlation is expected because of the larger measurement uncertainty for more distant events, which is due to their typically smaller S/N compared to events merging at lower redshifts.  In practice, we find that this procedure only leads to 0.6\% of the sample having nonphysical values $|\chi_\mathrm{eff}| > 1$ and 4.2\% of systems having nonphysical $z<0$, which we map back to $|\chi_\mathrm{eff}| = 1$ or $z=0$. We claim that this bias is small and does not affect our results as we still find that events with larger $\bar{z}$ have broader $\chi_\mathrm{eff}$ distributions. 

In Figure~\ref{fig:model_vs_GWTC3}, we show the comparison of the median $f^\mathrm{GWTC-3}_{\chi_\mathrm{eff}>\chi_0}(z)$ computed on the sample of mock GWTC-3 catalogs with the model prediction. The GWTC-3 quantity is independently measured for each mock catalog in the interval $z\in [0,1]$ by counting the events meeting the $\chi_\mathrm{eff}>\chi_0$ condition in the discrete redshift bin $\Delta z$ with constant cosmic time bin of size of $\Delta t = 1.6\,\mathrm{Gyr}$ and then quote the median value at each redshift bin. We then compare our model predictions with the inclusion of mock uncertainties by overlaying the median and 90\% CI $f^\mathrm{det}_{\chi_\mathrm{eff}>\chi_0}(z)$. 
We conclude that our model cannot be ruled out given the current GWTC-3 sample. 
Even though our model 90\% CI overlaps with the median inferred GWTC-3 value, a closer comparison with the model median indicates that our model slightly overpredicts the fraction of highly spinning BBHs. This could be due, e.g., to an overprediction of the fraction of highly spinning BBHs formed from the CHE channel which dominates over BBHs formed from the CE channel in the LIGO--Virgo detectable population (see Appendix~\ref{sec:no_channel}) or the existence of an additional channel contributing to the detectable BBH population with small BH spins \citep[see e.g., dynamical formation in globular clusters,][]{2021ApJ...910..152Z}. Other model uncertainties are discussed in Section~\ref{sec:discussion}.
%We also notice that for $f_{\chi_\mathrm{eff}>0.2}(z>0.6)$ the 90\% CI GWTC-3 inferred quantity manifests the largest possible uncertainty, namely $\Delta f^\mathrm{GWTC-3}_{\chi_\mathrm{eff}>\chi_0}(z>0.6) \in [0,1]$. For $\chi_\mathrm{eff}>0.5$ this level of uncertainty is reached for $z>0.8$.

We next infer the underlying fraction $f_{\chi_\mathrm{eff}>\chi_0}(z)$ by fitting a model for the astrophysical BBH population to the GWTC-3 data. We jointly fit the mass ($m_1$, $m_2$), spin $\chi_\mathrm{eff}$, and redshift $z$ distribution, allowing the $\chi_\mathrm{eff}$ distribution to evolve redshift but for simplicity neglecting possible correlations between other parameters:
\begin{equation}
    p_\mathrm{pop}(m_1, m_2, \chi_\mathrm{eff}, z) = p(m_1, m_2) p(\chi_\mathrm{eff} \mid z) p(z).
\end{equation}

For the mass distribution, $p(m_1, m_2)$, we use the Broken Power Law model from~\citet{2021ApJ...913L...7A} and for the redshift distribution, $p(z)$, we assume the merger rate evolves as a power law in $(1+z)$~\citep{2018ApJ...863L..41F}. We model the redshift-dependent spin distribution $p(\chi_\mathrm{eff} \mid z)$ as a mixture model between a ``zero-spin" component, approximated as a narrow Gaussian centered at $\chi_\mathrm{eff} = 0$ with standard deviation 0.03, and a ``positive spin'' component, for which we use a Gaussian distribution $\mathcal{N}^\mathrm{T}$ with mean $0.2 < \mu_p < 0.5$ and standard deviation $0.05 < \sigma_p < 0.5$ truncated to the range [0, 1] to reflect our model predictions. We take the mixture fraction $A$ between the zero and positive spin components to be a logistic function of $z$ (so that it is always within $0 < A < 1$), described by two free parameters, $A(z = 0)$ and $A(z = 1)$. We therefore have
\begin{equation}
\label{eq:pop-chieff-z}
    p(\chi_\mathrm{eff} \mid z) = \left(1 - A(z)\right)\mathcal{N}_{\mu = 0, \sigma = 0.03}(\chi_\mathrm{eff})
 + A(z)\mathcal{N}^\mathrm{T}(\chi_\mathrm{eff} \mid \mu_p, \sigma_p),
\end{equation}
where 
\begin{equation}
    A(z) = \left(1 + B\exp(k z)\right)^{-1},
\end{equation}
with $B = {A^{-1}(z = 0)} - 1$ and $k = \log \left(A^{-1}(z = 1) -1 \right) - \log(B)$.
We fit for all population parameters by sampling from a hierarchical Bayesian likelihood with PyMC3~\citep{pymc} \citep[see e.g.,][]{2019PASA...36...10T,2019MNRAS.486.1086M,2020arXiv200705579V}, using the GWTC-3 detector sensitivity estimates covering the first three observing runs and the parameter estimation samples for the GWTC-3 BBH events~\citep{LIGO-P1800370,LIGO-P2000223,10.5281_zenodo.5117703,ligo_scientific_collaboration_and_virgo_2021_5636816,ligo_scientific_collaboration_and_virgo_2021_5546663}. We use flat priors on all parameters within their ranges specified above.

\begin{figure}
\centering
\includegraphics[width = 0.5\textwidth]{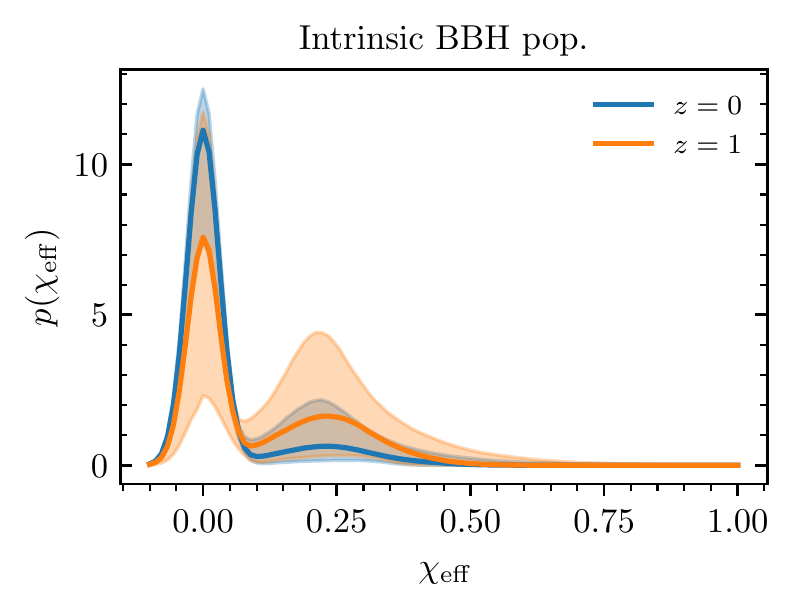}
\caption{Underlying $\chi_\mathrm{eff}$ distribution from fitting the population model of Eq.~\eqref{eq:pop-chieff-z} to the GWTC-3 BBH events. We plot the $\chi_\mathrm{eff}$ population distribution at two redshift slices, $z = 0$ (blue) and $z = 1$ (orange). Solid lines denote the median and shaded bands denote the 90\% CI.}
\label{fig:pop-chieff}
\end{figure}

\begin{figure}
\centering
\includegraphics[width = 0.5\textwidth]{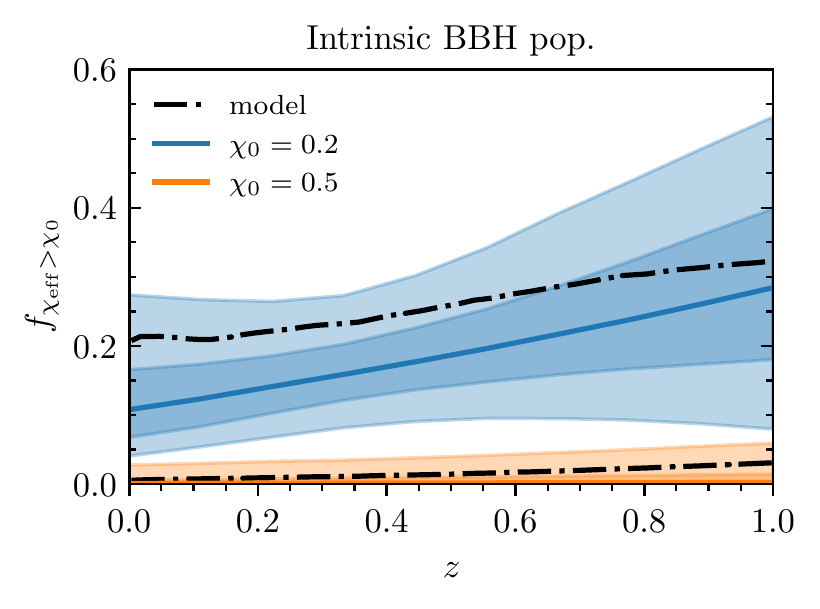}
\caption{High-$\chi_\mathrm{eff}$ fractions in the underlying distribution inferred from the population fit described in Section~\ref{sec:GWTC-3_corr}, in blue and orange according to the legend. Lighter contour colors indicate larger CIs of 50\% and 90\%, respectively. The fraction of BBH systems with large positive spins in the underlying population may increase with increasing redshift (credibility $82\%$), consistent with our model predictions (black). We do not yet have enough BBH events at $z \sim 1$ to accurately measure the $\chi_\mathrm{eff}$ distribution at high $z$ and therefore cannot confidently conclude that the distribution is evolving.}
\label{fig:underlying-fchieff-GWTC3}
\end{figure}

\begin{figure}
\centering
\includegraphics[width = 0.5\textwidth]{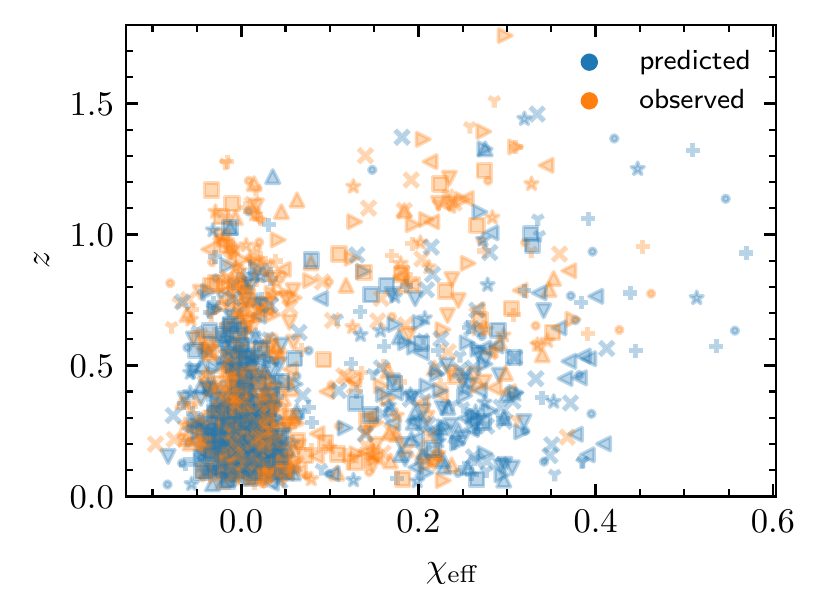}
\caption{Redshift and effective spin parameters of the 69 confident BBH observations drawn from the GWTC-3 posteriors (orange; ``observed'') compared to 69 draws from the inferred distribution fit (blue; ``predicted'') described in Section~\ref{sec:GWTC-3_corr}. Each marker shape corresponds to a different set of 69 draws. We plot ten total sets. The inferred model sometimes over-predicts the largest observed $\chi_\mathrm{eff}$, while the bulk of both observed and predicted draws cover an equivalent portion of the $z-\chi_\mathrm{eff}$ plane in a comparable abundance, confirming that the inferred model is a good fit for the data.}
\label{fig:ppc-chieff-z}
\end{figure}

The inferred intrinsic $\chi_\mathrm{eff}$ population distribution at two redshifts, $z = 0$ and $z = 1$, is shown in Figure~\ref{fig:pop-chieff}. At $z = 0$ the positive-spin component is constrained to be small, whereas at $z = 1$, the data permit a larger fraction of systems with high $\chi_\mathrm{eff}$, although the overall constraints are more uncertain and more closely resemble the prior. We can directly compare the underlying high-$\chi_\mathrm{eff}$ fractions $f_{\chi_\mathrm{eff} > 0.2}$ and $f_{\chi_\mathrm{eff} > 0.5}$ inferred under this fit to the low-redshift $z < 1$ predictions in the leftmost panel of Figure~\ref{fig:model_f_chi}. In Figure~\ref{fig:underlying-fchieff-GWTC3}, we show the inferred intrinsic $f_{\chi_\mathrm{eff} > 0.2}(z)$ and $f_{\chi_\mathrm{eff} > 0.5}(z)$ versus our astrophysical model predictions. The intrinsic fractions are broadly consistent with the model predictions, although the data prefer slightly smaller fractions of large positive $\chi_\mathrm{eff}$ at all $z$, similar to the conclusions of Figure~\ref{fig:model_vs_GWTC3} regarding the observed fractions. To reiterate, this could be due, e.g., to an overprediction of the contribution of the CHE channel (see Appendix~\ref{sec:no_channel}) or the non-negligible contribution of an additional channel with small BH spins. Other model uncertainties are discussed in Section~\ref{sec:discussion}.

We verify the goodness-of-fit of the inferred model by performing posterior predictive checks. Figure~\ref{fig:ppc-chieff-z} shows the comparison between the $(z,\chi_\mathrm{eff})$ parameters of ten mock GWTC-3 catalogs versus ten sets of 69 events drawn from the inferred model. This test is analogous to the posterior predictive check in Figure 2 of~\citet{2021ApJ...912...98F}. Each of the ten sets (plotted with a different marker size) corresponds to one draw from the inferred population hyperposterior. We reweight the single-event posterior from each GWTC-3 event to the population distribution specified by the hyperposterior draw, and draw one $(z,\chi_\mathrm{eff})$ sample per event. We then draw a set of 69 predicted events from the same population distribution, conditioned on detection. We can see that the inferred model sometimes over-predicts the largest observed $\chi_\mathrm{eff}$, while the bulk of both the observed and predicted draws cover an equivalent portion of the $z-\chi_\mathrm{eff}$ plane in a comparable abundance, confirming that the inferred model is a good fit to the data.

Despite the suggestive hint that $f_{\chi_\mathrm{eff} > 0.2}(z)$ increases with $z$, we are not yet able to confidently identify that the $\chi_\mathrm{eff}$ distribution varies with redshift under our parameterization. More precisely, we constrain $f_{\chi_\mathrm{eff} > 0.2}(z=0.3) > 0.06$ at 99\% credibility and find that $f_{\chi_\mathrm{eff} > 0.2}(z)$ increases with increasing redshift at 82\% credibility.
Our conclusions are consistent with the results of \citet{2022arXiv220401578B}, who find that the width of the $\chi_\mathrm{eff}$ distribution likely broadens with increasing redshift, but do not find compelling evidence that the mean $\chi_\mathrm{eff}$ evolves with redshift.\footnote{Our parameterization for the $\chi_\mathrm{eff}$ distribution is most similar to the model \citet{2022arXiv220401578B} consider in their Section 4.3 with the ``Prior 3'' variation. \citet{2022arXiv220401578B} analysis consider alternative models for the parameterization of the redshift evolving $\chi_\mathrm{eff}$ distribution other than the one assumed here, still reaching similar conclusions.} Our model for field BBH formation predicts that the mean of the $\chi_\mathrm{eff}$ distribution must increase with redshift. 

We further caution that our phenomenological population fit makes the simplifying assumption that the mass distribution is independent of spin and redshift, despite the fact that we predict a correlation between total mass, $\chi_\mathrm{eff}$, and redshift. However, given current statistical uncertainties on the inferred intrinsic $f_{\chi_\mathrm{eff} > \chi_0}(z)$, we do not expect our systematic errors on this inferred quantity from mismodeling the population distribution to be significant. However, with future data it will be important to allow for $\chi_\mathrm{eff}$ to vary with both mass and redshift in BBH population fits, because as Figures.~\ref{fig:chi_eff_model} and~\ref{fig:model_f_chi} show, some of the observed $\chi_\mathrm{eff}$ evolution in the LIGO-Virgo catalog will be due to an underlying correlation between $\chi_\mathrm{eff}$ and mass. These conclusions are corroborated by \citet{2022arXiv220401578B}, who find that the preference for $\chi_\mathrm{eff}$ to correlate with redshift is stronger than a possible correlation with primary mass, although the two scenarios can be confused for each other. 
 
\section{Discussion}\label{sec:discussion}

In this work, we considered a fiducial model for isolated binary evolution. However, model uncertainties can potentially alter BBH observable distributions and rates (see e.g. \citealt{2021arXiv211205763B} for an extended overview of such uncertainties). Here, we are interested in astrophysical uncertainties which may alter the $\chi_\mathrm{eff}-z$ joint distribution. 

Our fiducial model assumes efficient angular momentum transport inside stars which leads to the formation of non-spinning first-born BHs for the CE and SMT channels. Alternatively, a less efficient angular momentum transport would lead to non-negligible birth spins (see e.g. some model variations in \citealt{2020A&A...636A.104B}) which would consequently raise our estimated fraction of highly spinning BBHs $f_{\chi_\mathrm{eff}>\chi_0}(z)$ as the CE and SMT channels dominate the intrinsic BBH population. Nevertheless, current observations are consistent with low birth spins of $\lesssim 0.1$  for isolated BHs \citep{2021ApJ...913L...7A,2021arXiv211103634T,2020ApJ...895..128M,2021ApJ...910..152Z}.

In \citet{2021A&A...647A.153B}, the impact of mass-transfer physics uncertainties on the $\chi_\mathrm{eff}$ distribution of BBHs formed from the CE and SMT channels was investigated, accounting for uncertainties in (i) the unknown efficiency of CE ejection in the $\alpha_\mathrm{CE}-\lambda$ parametrization \citep[see, e.g.,][for a review]{2013A&ARv..21...59I}, (ii) the SMT accretion efficiency onto BHs, and (iii) the criteria for mass-transfer stability. The first uncertainty directly impacts the relative fraction of highly rotating BBHs in the CE channel as the $\alpha_\mathrm{CE}$ parameter approximately linearly scales with the orbital separation post CE. For a wide range of $\alpha_\mathrm{CE}\in[0.2,5]$, \citet{2021A&A...647A.153B} showed that the BBH fraction of systems with $\chi_\mathrm{eff}> 0.1$ formed from the CE channel can vary from 0.54 to 0.82 where the merger rate density might also vary by up to one order of magnitude. Nevertheless, \citet{2021A&A...647A.153B} showed how both $\alpha_\mathrm{CE}$ extremes include a non-zero fraction of tidally spun-up BBHs in the CE channel.
The second uncertainty affects the initially negligible spin of the first-born BH of a BBH systems formed through the SMT channel. In the case of highly super-Eddington accretion efficiency onto BHs, \citet{2021A&A...647A.153B} showed how a non-negligible fraction of first-born BHs could be spun up due to accretion. However, in such cases, depending on the super-Eddington accretion efficiency, \citet{2021A&A...647A.153B} found a suppression of the SMT merger rate density up to two orders of magnitude. This occurs because conservative mass transfer is less efficient than unconservative mass transfer in leading to tight BH-WR systems, leading to less BBH systems that can merge in a Hubble time. Finally, the third uncertainty directly impacts the relative fraction of systems that undergo either stable or unstable mass transfer and, hence, CE or SMT evolution. We now examine how these uncertainties might affect the presented $\chi_\mathrm{eff}-z$ correlation.

In the present study, we assumed inefficient CE ejection, namely the model with $\alpha_\mathrm{CE}=0.5$ of \citep{2021A&A...647A.153B}. A smaller value than what was assumed here would lead to a more significant fraction of tidal spun-up BBHs. Because the CE channel dominates the intrinsic BBH population, such a scenario would increase the predicted quantity $f_{\chi_\mathrm{eff}>\chi_0}(z)$. In contrast, a more efficient assumption for CE ejection would lead to a smaller fraction of systems that are tidally spun up. For the detectable fraction $f^\mathrm{det}_{\chi_\mathrm{eff}>\chi_0}(z)$, we expect a small impact of this assumption as the detectable population of BBHs is dominated by the SMT and CHE channels. 
Since \citet{2021A&A...647A.153B} showed that at $\alpha_\mathrm{CE}=5$ there is still a fraction of highly rotating BBHs formed from the CE channel with a median  $\bar{\chi}^\mathrm{CE}_\mathrm{eff}=0.16$, we can claim that our model will always display a monotonically increasing $f_{\chi_\mathrm{eff}>\chi_0}(z)$ regardless of the $\alpha_\mathrm{CE}$ value in the CE parameterization. 

Our fiducial model assumed Eddington limited mass-transfer accretion efficiency onto BHs. A super-Eddington accretion efficiency onto BHs would boost the fraction of highly spinning BBHs formed from the SMT channel, and, hence, positively contribute to larger values of $f_{\chi_\mathrm{eff}>\chi_0}(z)$ and $f^\mathrm{det}_{\chi_\mathrm{eff}>\chi_0}(z)$. However, given that the BBH merger rate from the SMT channel (both detected and intrinsic) drops by up to two orders of magnitude compared to our fiducial model when increasing the allowed accretion rate onto BHs \citep[see Table~1 of][]{2021A&A...647A.153B}, we would expect a smaller intrinsic contribution to the SMT channel than the one modeled here. 

Both uncertainties (i) and (iii) might lead to a smaller relative contribution of the CE channel to the total BBH population than what is assumed here, where the CE channel dominates the $f_{\chi_\mathrm{eff}>\chi_0}(z)$ behavior. Moreover, recent studies employing detailed binary simulations point towards an overestimation of systems evolving through and surviving CEs due to envelope stripping during the CE ceasing earlier than what is assumed in rapid population synthesis codes \citep{2019ApJ...883L..45F,2019A&A...628A..19Q,2021A&A...645A..54K,2021A&A...650A.107M,2021ApJ...922..110G}. Therefore, it is natural to ask ourselves what would happen to the modeled $f_{\chi_\mathrm{eff}>\chi_0}(z)$ fraction if the CE channel is negligible compared to SMT and CHE. In such a scenario, given our model, one would expect that most binaries evolving through CE would either evolve through SMT or successfully emerge from the CE at wider orbital separations. In the first case this would lead to a SMT contribution that is similar or greater than what is modeled here. The second case would lead to a reduced fraction of tidally spun-up CE systems, similarly to the outcome of choosing an efficient $\alpha_\mathrm{CE}$ values. If the remaining SMT and CHE channels retain a similar fraction of highly spinning BBHs to what is modeled here, one would find $f_{\chi_\mathrm{eff}>0.2}(z<4)\simeq 0.2$ which would eventually decay at larger redshifts while the LIGO--Virgo detectable population would still exhibit a monotonically increasing behaviour since the contribution of CE systems to the LIGO--Virgo detectable population is small ($f^\mathrm{det}_\mathrm{CE}(z>0.25)<10\%$). In Appendix~\ref{sec:no_channel}, we show how Fig.~\ref{fig:model_f_chi} would change given the omission of the CE channel from our fiducial model. At low redshifts, $z<1$, we find that the intrinsic fraction $f_{\chi_\mathrm{eff}>0.2}$ of this alternative model is still consistent with the 90\% CI of GWTC-3 constraints shown in Figure~\ref{fig:underlying-fchieff-GWTC3}.

Similar to this last point, we alternatively entertain the idea of what would happen to the $\chi_\mathrm{eff}-z$ correlation if either the SMT or CHE channels contributions are negligible. This might happen, for example, if we overestimate the contribution of the initial conditions parameter space at low orbital periods that leads to SMT or CHE evolution. In Appendix~\ref{sec:no_channel}, we show that the presented correlation would still be observed. In both cases we still recover monotonically increasing fractions $f_{\chi_\mathrm{eff}>\chi_0}(z)$ and $f^\mathrm{det}_{\chi_\mathrm{eff}>\chi_0}(z)$. However, we note that the model with the omission of the SMT channel manifests a larger, relatively constant $f_{\chi_\mathrm{eff}>0.2}(z<1)\simeq 0.45$ which is inconsistent with the 90\% CI of GWTC-3 constraints in Figure~\ref{fig:underlying-fchieff-GWTC3} of $f_{\chi_\mathrm{eff}>0.2}(z<0.4)<0.3$. Finally, we find that the model that excludes the CHE channel has both an intrinsic $f_{\chi_\mathrm{eff}>\chi_0}$ and LIGO--Virgo detectable $f^\mathrm{det}_{\chi_\mathrm{eff}>\chi_0}$ closer to the median GWTC-3 inferred constrains of Figures~\ref{fig:model_vs_GWTC3} and \ref{fig:underlying-fchieff-GWTC3}.

\section{Conclusions}\label{sec:conclusions}

In this paper, we investigated the $\chi_\mathrm{eff}-z$ correlation of field-formed merging BBHs. An increasing fraction of highly spinning BBHs as a function of redshift is expected. At higher redshifts, stars are formed at lower metallicities, experience weaker stellar wind mass loss, and consequently can maintain their short orbital separations and undergo tidal spin up. We quantified this correlation by the fraction of systems with $\chi_\mathrm{eff}>\chi_0$ as a function of redshift, $f_{\chi_\mathrm{eff}>\chi_0}(z)$. For our fiducial model of field BBHs, which includes the potential contribution of CE, SMT, and CHE channels, this quantity for $\chi_0\in[0.2,0.5]$ shows a monotonically increasing behavior as a function of redshift in the underlying BBH population. We also presented predictions for the detectable $f^\mathrm{det}_{\chi_\mathrm{eff}>\chi_0}(z)$ for the LIGO--Virgo detector network at O3 sensitivity and the Einstein Telescope. Because of the smaller horizons of current GW detectors ($z\simeq 1$), the origin of the monotonically increasing LIGO--Virgo $f^\mathrm{det}_{\chi_\mathrm{eff}>\chi_0}(z)$ quantity is different than the intrinsic BBH population or that which the Einstein Telescope will observe in the future. Such differences originate from different BH mass distributions of the various channels. On average, highly rotating BBHs formed from the CHE channel are more massive than tidally spun up BBH systems formed from the CE channel. Hence, LIGO--Virgo detector selection effects favour high BH masses and lead to different observational horizons for different channels. We find that, in contrast to the intrinsic distribution where the $\chi_\mathrm{eff}-z$ correlation is dominated by tidal spun-up BBHs from the CE channel, the CHE channel dominates the LIGO--Virgo detected $\chi_\mathrm{eff}-z$ correlation above $z>0.4$.

Finally, assuming isolated binary evolution dominates the detected population of merging BBHs, we performed a model comparison between our fiducial model and LIGO--Virgo GWTC-3 data. We find that current observations favor the prediction of our model that there is a positive correlation between $\chi_\mathrm{eff}$ and $z$. 
Such a conclusion is consistent with the results of \citet{2022arXiv220401578B} who found that the width of the $\chi_\mathrm{eff}$ distribution likely broadens with increasing redshift, event though they did not find compelling evidence in favor of a redshift evolving mean $\chi_\mathrm{eff}$. 
Additionally, our model prediction at low redshifts of a large zero-spin BBH population with an additional subpopulation of systems with spin vectors preferentially aligned to the orbital angular momentum is in agreement with \citet{2021PhRvD.104h3010R} and \citet{2021ApJ...921L..15G} reanalysis of GWTC-2 events.
Moreover, our results are consistent with the findings that investigated field BBH observable properties and rates \citep{2020A&A...635A..97B,2021A&A...647A.153B}, multi-channel model selection with GWTC-2 data \citep{2021ApJ...910..152Z}, potential constraints from LGRBs \citep{2022A&A...657L...8B}, and the current upper limits of the stochastic GW background \citep{2021arXiv210905836B}.
    
Considering future 3G GW detector facilities, we demonstrated that if isolated binary evolution plays a dominant role in the formation of merging BBHs in the Universe, 3G GW detectors will observe more of the merging BBHs in the Universe and a $\chi_\mathrm{eff} - z$ correlation that is more indicative of the behavior of the underlying population.

\begin{acknowledgements}
We thank Sylvia Biscoveanu and Christopher Berry for useful comments on this manuscript. This work was supported by the Swiss National Science Foundation Professorship grant (project number PP00P2\_176868). MF and MZ are supported by NASA through NASA Hubble Fellowship grants HST-HF2-51455.001-A and HST-HF2-51474.001-A awarded by the Space Telescope Science Institute, which is operated by the Association of Universities for Research in Astronomy, Incorporated, under NASA contract NAS5-26555. EZ acknowledges funding support from the European Research Council (ERC) under the European Union’s Horizon 2020 research and innovation programme (Grant agreement No. 772086). This material is based upon work supported by NSF's LIGO Laboratory which is a major facility fully funded by the National Science Foundation. All figures were made with the open-source Python module \texttt{Matplotlib} \citep{Hunter:2007}. This research made use of the python modules \texttt{Astropy} \citep{price2018astropy}, \texttt{iPhyton} \citep{PER-GRA:2007}, \texttt{Numpy} \citep{harris2020array} and \texttt{SciPy} \citep{2020SciPy-NMeth}.
\end{acknowledgements}

\bibliography{aanda}

\appendix

\section{Angular momentum loss due to pulsational pair-instability supernovae}\label{sec:AM_loss}

Mass loss due to PPISNe can play a role in depleting the angular-momentum reservoir of a collapsing star. Because the pulsations carry away the outer layers of the stars that carry most of the angular-momentum content of the star, this phenomena could have a major impact in reducing the spins of massive BHs. 

The impact of PPISNe on the spin of the second-born BH of tidally spun up BH-WR systems was briefly discussed in \citet{2022arXiv220302515Z}. For tidally spun-up systems with orbital periods $p<1\,\mathrm{day}$ and WR stellar masses of $M_\mathrm{WR}>40\,M_\odot$ at carbon depletion, the first panel of Figure~1 in \citet{2021RNAAS...5..127B} shows a small suppression of the second-born BH spin obtained from the WR stellar profile collapse of \texttt{MESA} BH-WR simulations from \citet{2021A&A...647A.153B}. Because WR stellar wind rates scale as a function of metallicity \citep{2001A&A...369..574V}, only binaries born at low metallicities (prevalently formed at high redshifts) will evolve to have WR stars in such a mass regime. Hence, for the CE channel, we expect this phenomena to have a small impact as on average the channel operates at smaller WR stellar mass. For the SMT channel, we find that in practice this phenomena is relevant only at large redshifts as this channel leads on average to more massive BH-WR star systems compared to the CE channel, resulting in a $f^\mathrm{SMT}_{\chi_\mathrm{eff}>0.2}(z\geq 7) \simeq 0.45$ plateau in Figure~\ref{fig:model_f_chi}. 

\begin{figure}
\centering
\includegraphics{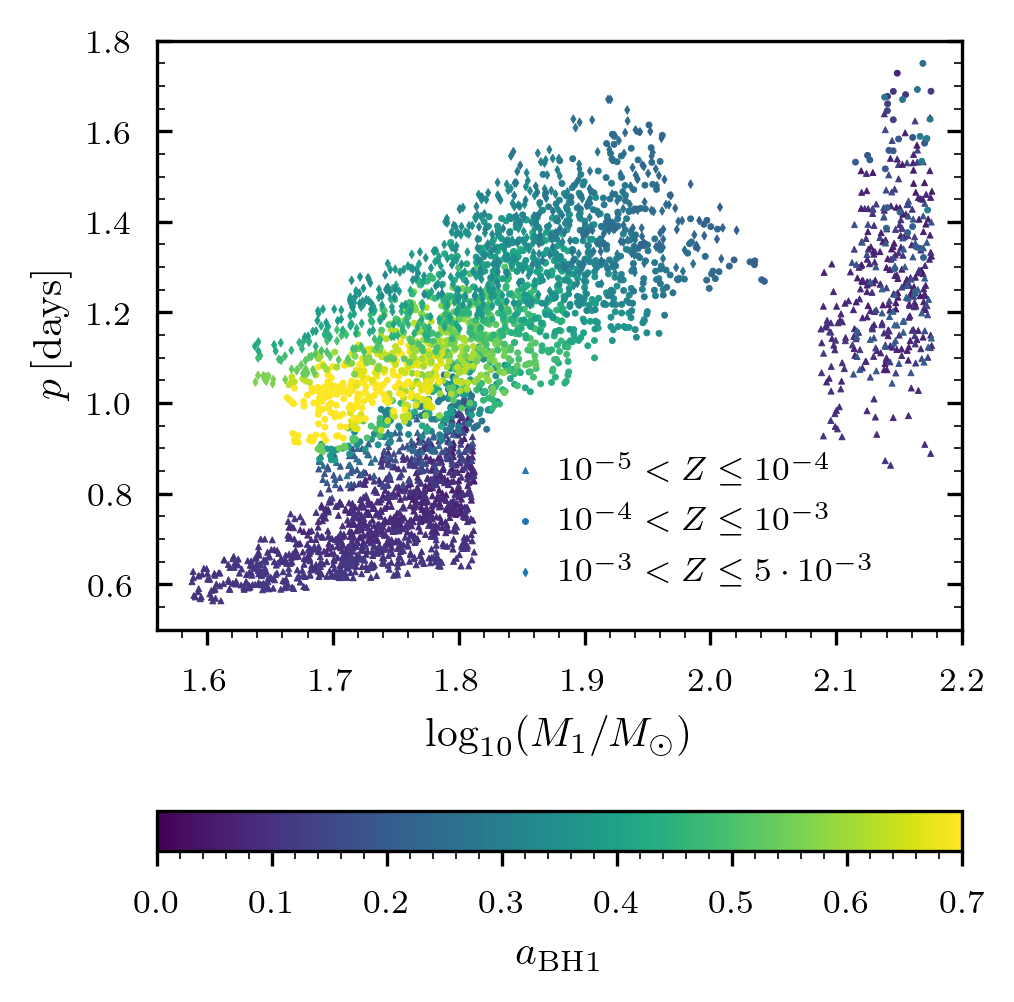}
\caption{Distribution of ZAMS binary orbital period, p, primary mass, $M_1$, and the final primary BH spin of systems evolving thorough the CHE channel to become merging BBHs in our fiducial model. In this sample we only include BBH systems with inspiral times less than the age of the Universe. Different markers differentiate metallicity regimes according to the legend. For visualisation purposes, we capped the color bar at $a_\mathrm{BH1}=0.7$ even though there are BHs approaching the general relativistic limit $a_\mathrm{BH1}=1$. Though binaries with $p<1 \,\mathrm{day}$ do tidally spin up and evolve through CHE, they later undergo mass loss due to PPISN which depletes the WR star of its angular momentum reservoir.}
\label{fig:CHE_BBH}
\end{figure}

In contrast, we find that the impact of PPISNe onto the spin of BHs formed from the CHE channel is not negligible as this channel only operates at low metallicities ($Z<5\cdot 10^{-3}$) and for massive stars. For metallicities $Z\leq10^{-4}$ the entire sample of merging BBHs evolving through the CHE channel is formed by stars with ZAMS primary masses $40\,M_\odot \lesssim M_1 \lesssim 70\,M_\odot$ which undergo PPISN. This occurs because at these low metallicities stellar wind mass loss is weaker compared to larger metallicities, and the stars reach the mass regime of PPISN, see Figure~A1 of \citet{2020MNRAS.499.5941D}. We note that in our fiducial model we do not simulate BBH formation above ZAMS primary masses of $150\,M_\odot$, hence Figure~A1 of \citet{2020MNRAS.499.5941D} should be read accordingly. On the other hand, the $10^{-4} < Z \leq 5 \cdot 10^{-3}$ parameter space leading to the formation of merging BBHs allows for direct collapse and, hence, conservation of angular momentum during the stellar profile collapse (with the exception of extremely highly rotating stars inducing disk formation). In Figure~\ref{fig:CHE_BBH}, we show the ZAMS binary conditions leading to merging BBH formation through the CHE channel, showing their final primary BH spins as a function of ZAMS initial orbital period and primary mass which can be directly compared to Figure~A1 of \citet{2020MNRAS.499.5941D}. We can see that for $Z\leq10^{-4}$ and ZAMS primary masses $\lesssim 70\,M_\odot$ the entire population of BBHs is composed of BBH systems with negligible spins as they have lost their high stellar angular momentum due to PPISN mass ejection.
The gap in the parameter space at $1.8 \lesssim \log_{10}(M_1/M_\odot) \lesssim 2.1$ for $Z\leq10^{-4}$ binaries in Figure~\ref{fig:CHE_BBH} is due to pair-instability supernovae leaving no remnant. For binaries with $Z > 10^{-4}$, this portion of the parameter space is present at larger ZAMS primary masses and orbital periods \citep[see Figure~A1 of][]{2020MNRAS.499.5941D}.
The impact of PPISN onto the BH spin of BBHs formed from the CHE channel at extremely low metallicities explains the monotonically decreasing behaviour of $f^\mathrm{CHE}_{\chi_\mathrm{eff}>\chi_0}(z)$ as a function of increasing redshift as the Universe forms more stars at these low metallicities.

\section{The $\chi_\mathrm{eff}-z$ correlation with channel exclusion}\label{sec:no_channel}

\begin{figure*}
\centering
\includegraphics{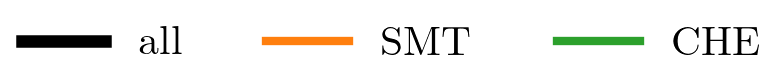}\\
\includegraphics{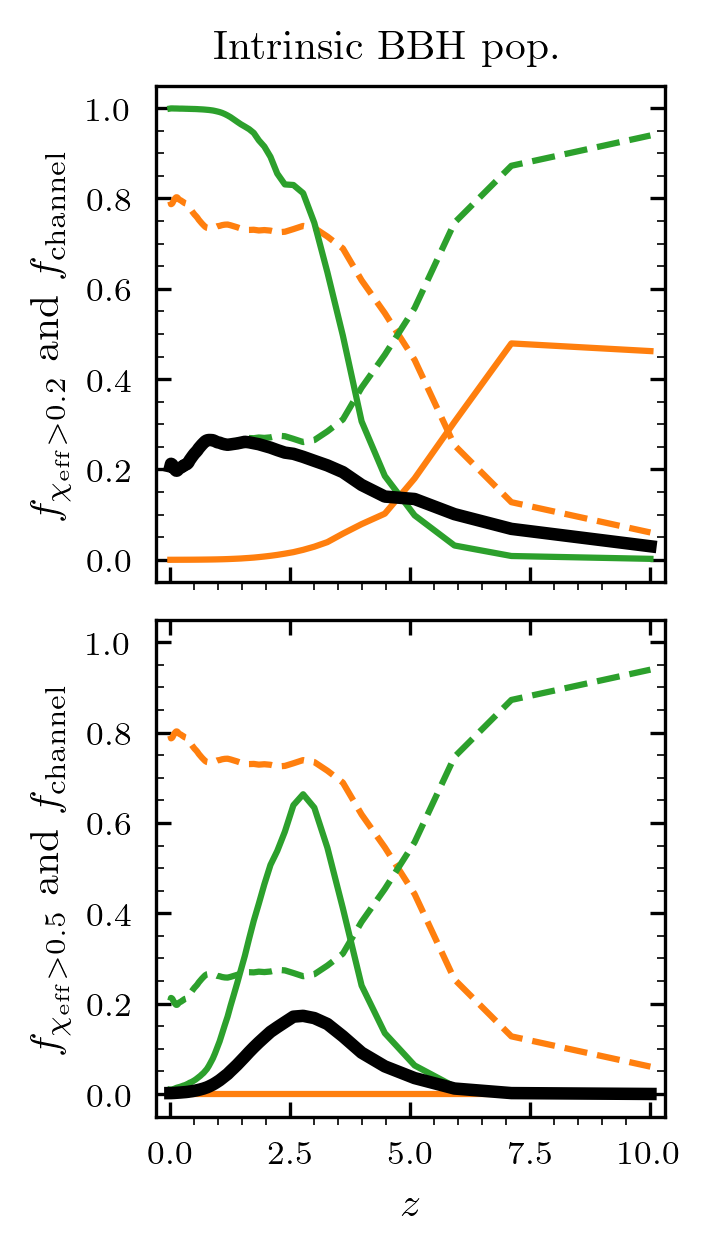} %[width=0.495\linewidth]
\includegraphics{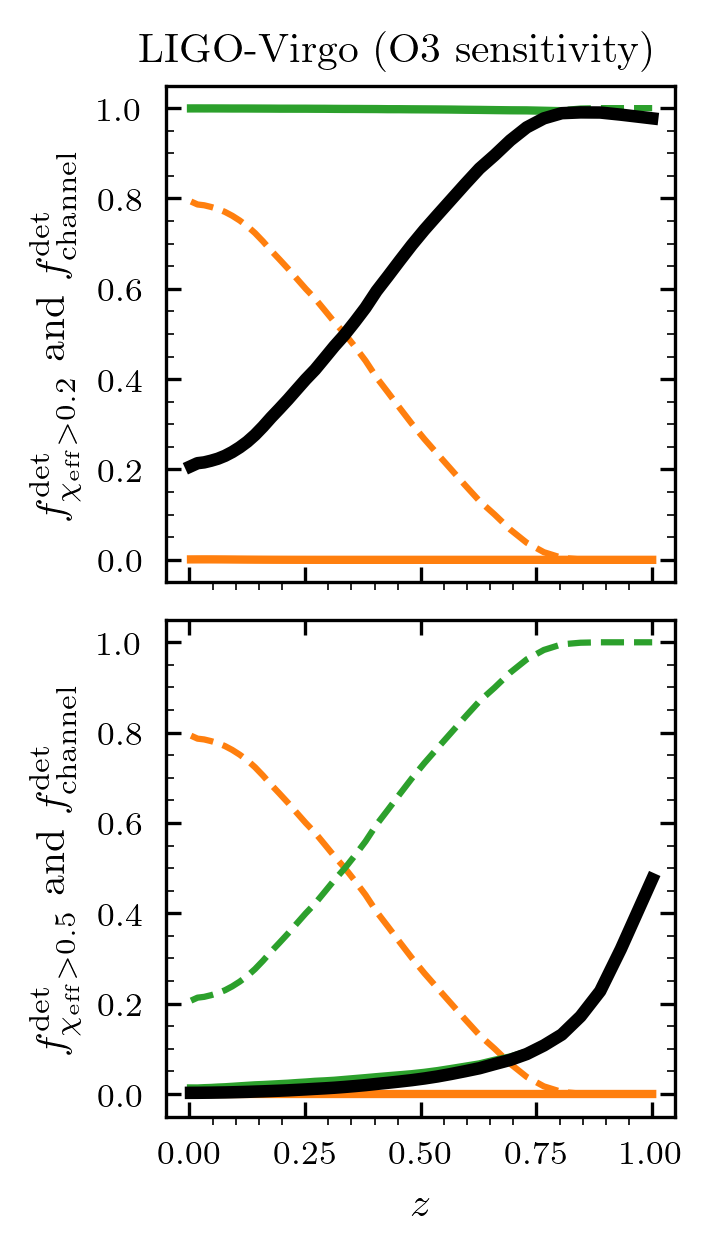}
\includegraphics{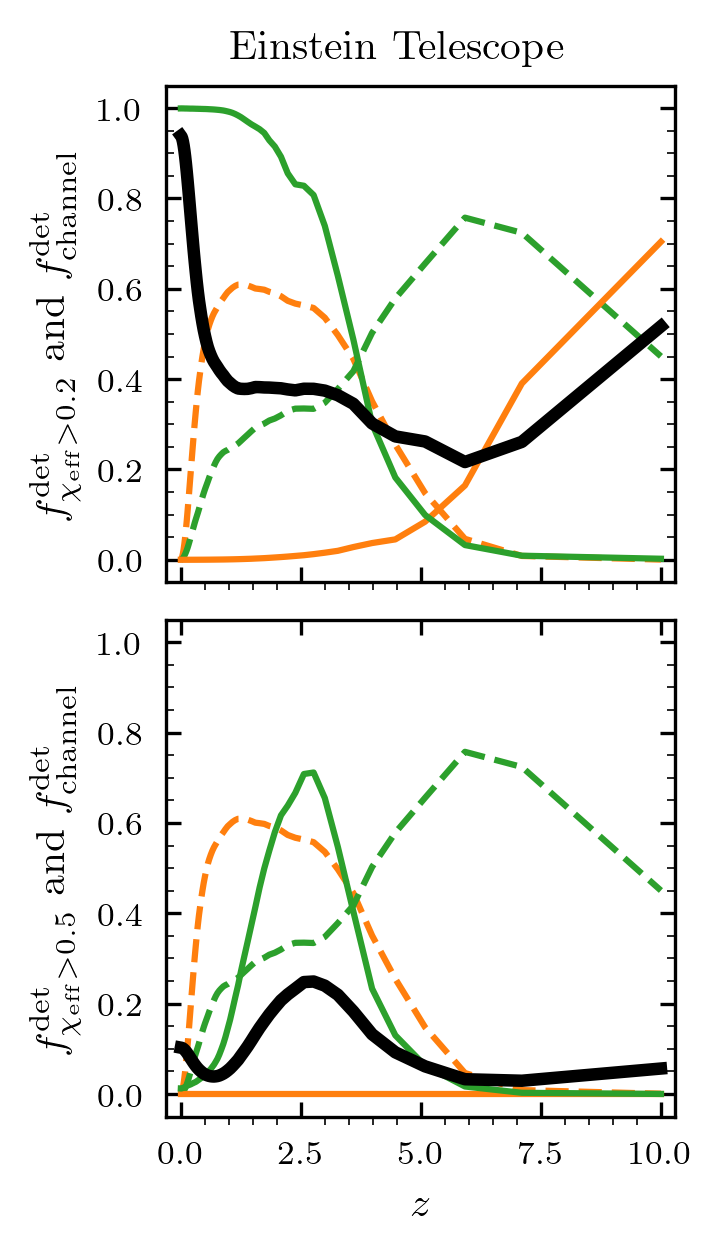}
\caption{Same as Figure~\ref{fig:model_f_chi} but the model of isolated binary evolution excludes the CE channel.}
\label{fig:model_f_chi_without_CE}
\end{figure*}

In this appendix section we show the impact to our results presented in Figure~\ref{fig:model_f_chi} in the hypothetical scenario that one of the three channels considered has a negligible contribution to the formation of merging BBHs.

First, let us consider neglecting the CE channel. Factors that might lead to this hypothetical scenario are discussed in Section~\ref{sec:discussion}. Figure~\ref{fig:model_f_chi_without_CE} shows how the results presented in Figure~\ref{fig:model_f_chi} would change under this assumption. In this alternative model, the intrinsic fraction $f_{\chi_\mathrm{eff} > 0.2}(z)$ is mainly supported by highly spinning BBHs formed from the CHE channel at $z<5$, while at larger redshift the SMT channel contributes with a larger fraction of tidally spun-up BHs. However, we notice that in contrast to our fiducial model the intrinsic fraction $f_{\chi_\mathrm{eff} > 0.2}(z)$ is monotonically decreasing. On the other hand the LIGO--Virgo detectable BBH population shows a similar behaviour as the fiducial model. This occurs as the CE channel contribution to the LIGO--Virgo detectable population is small compared to the SMT and CHE channels, since the CE channel leads to less massive BBHs (cf. Figure~\ref{fig:model_f_chi}).

\begin{figure*}
\centering
\includegraphics{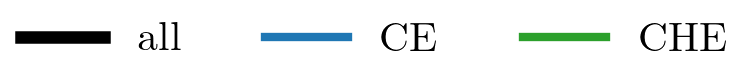}\\
\includegraphics{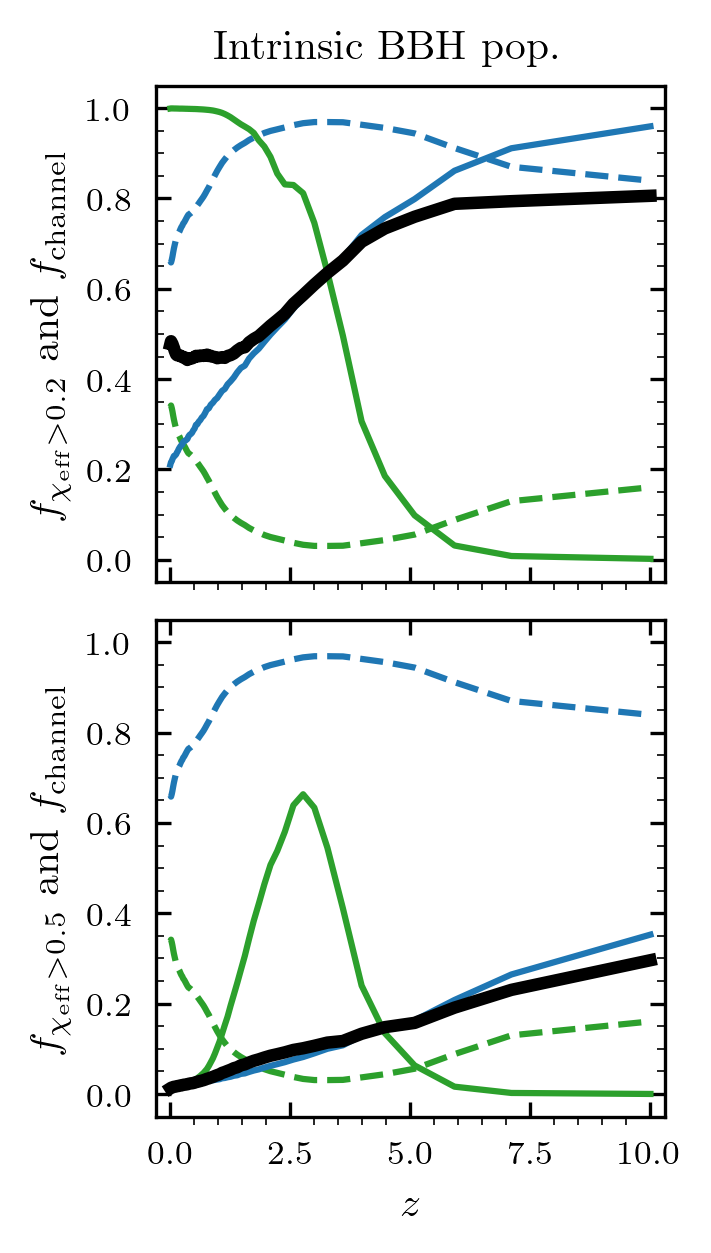} %[width=0.495\linewidth]
\includegraphics{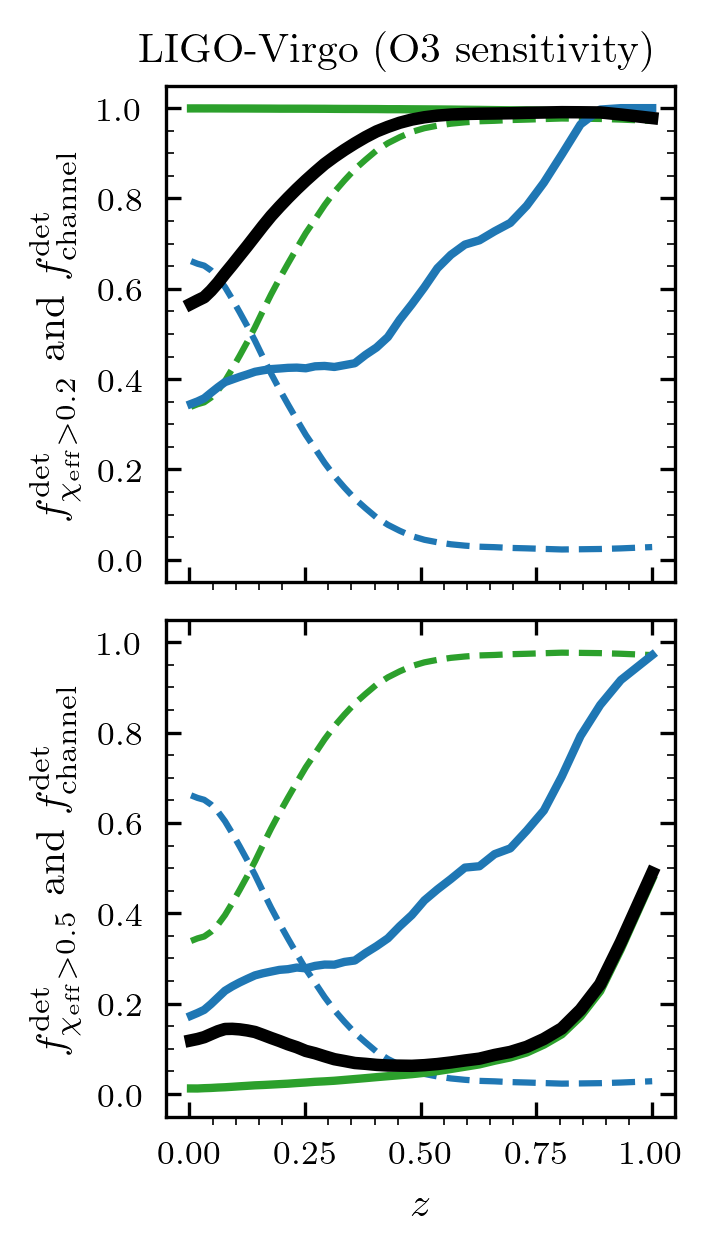}
\includegraphics{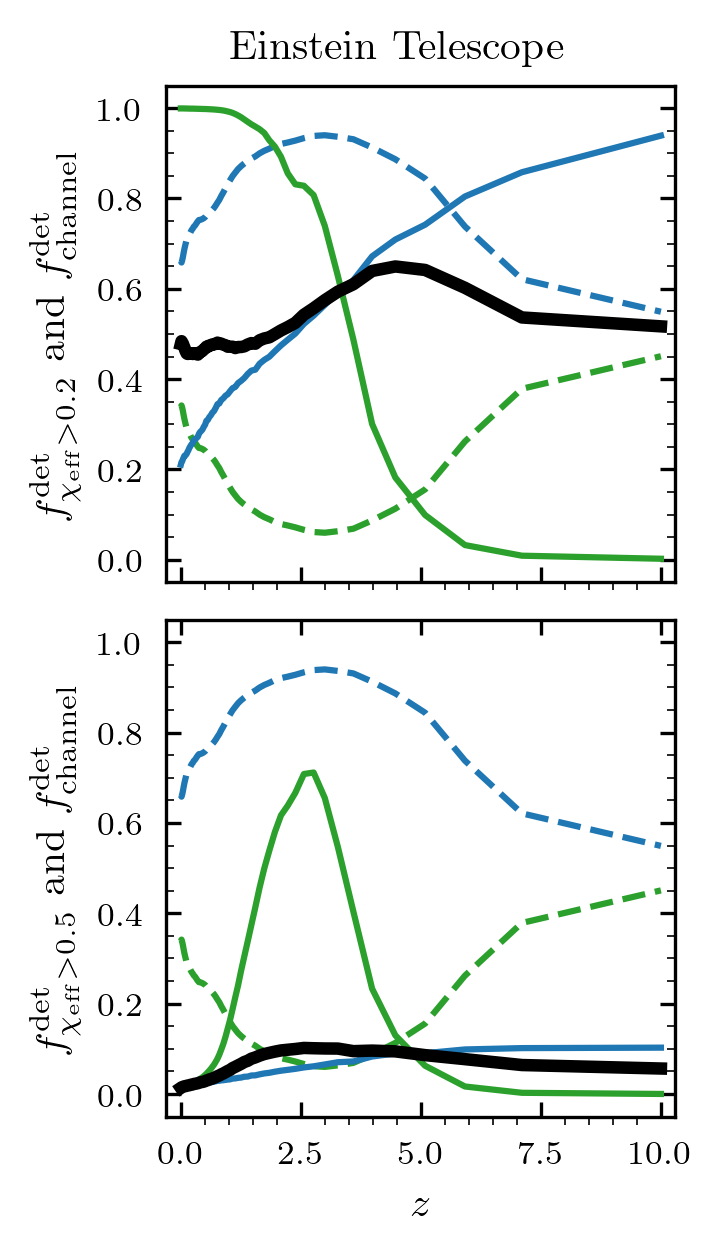}
\caption{Same as Figure~\ref{fig:model_f_chi} but the model of isolated binary evolution excludes the SMT channel.}
\label{fig:model_f_chi_without_SMT}
\end{figure*}

Second, let us consider neglecting the SMT channel. Figure~\ref{fig:model_f_chi_without_SMT} shows how the results presented in Figure~\ref{fig:model_f_chi} would change under this assumption. Because at low redshifts ($z<5$) the SMT channel mostly contributes to the intrinsic distribution with non-spinning BBHs, this alternative scenario leads to a larger $f_{\chi_\mathrm{eff} > 0.2}(z)$ fraction compared to the fiducial model. This hypothetical scenario would result in a LIGO--Virgo detectable BBH population $f^\mathrm{det}_{\chi_\mathrm{eff} > 0.2}(z)\gtrsim 0.6$, in tension with GWTC-3 observations. 

\begin{figure*}
\centering
\includegraphics{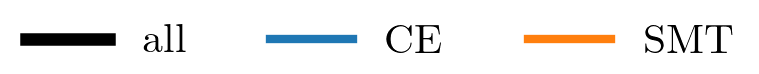}\\
\includegraphics{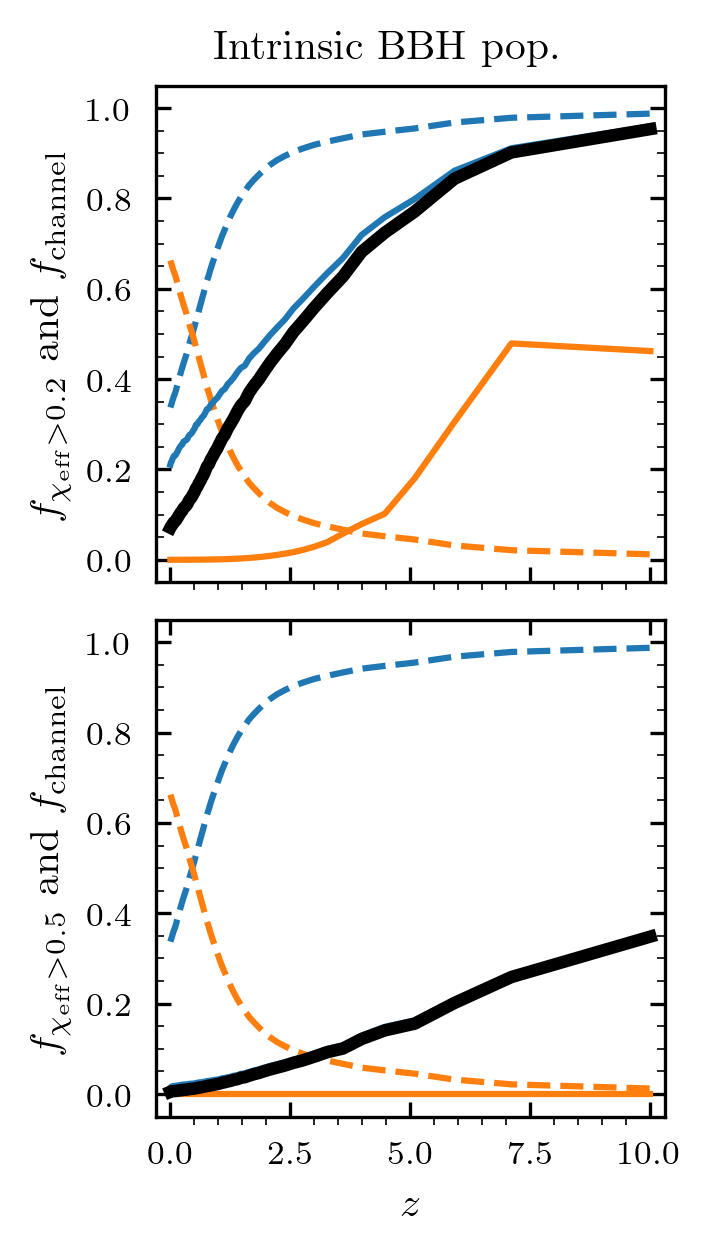} %[width=0.495\linewidth]
\includegraphics{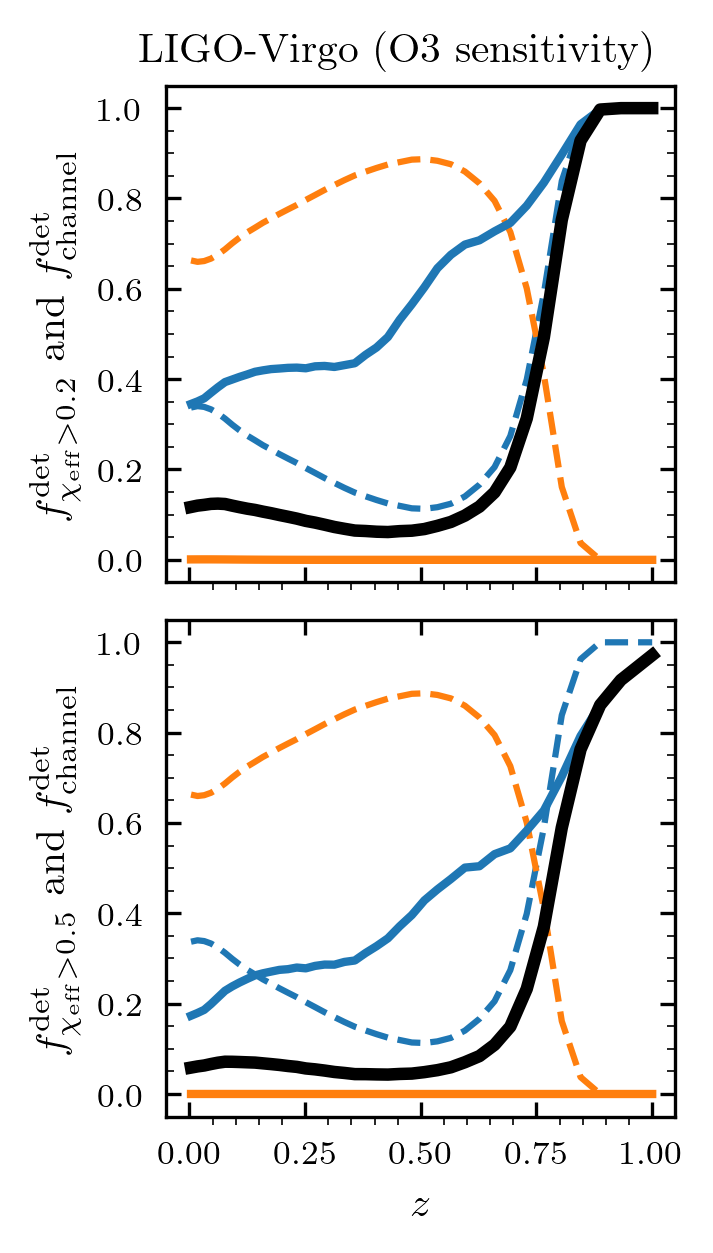}
\includegraphics{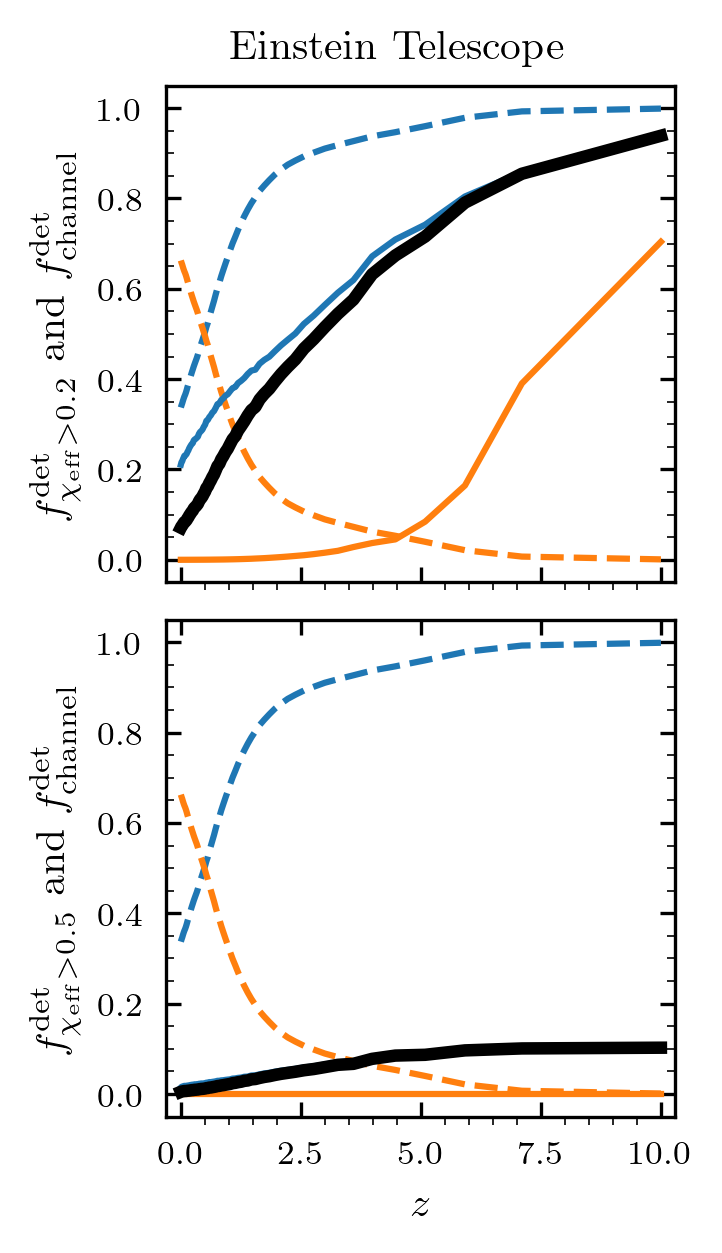}
\caption{Same as Figure~\ref{fig:model_f_chi} but the model of isolated binary evolution excludes the CHE channel.}
\label{fig:model_f_chi_without_CHE}
\end{figure*}

Last, let us consider neglecting the CHE channel.
As discussed in Section~\ref{sec:discussion} this might occur, for example, in the hypothetical case where the abundance of binary stars at ZAMS with short orbital periods ($p<2\,\mathrm{days}$) is overestimated. This alternative model is presented in Figure~\ref{fig:model_f_chi_without_CHE}. We can see that the $f_{\chi_\mathrm{eff} > \chi_0}(z)$ distribution is similar to what is presented in Figure~\ref{fig:model_f_chi}. This is explained by the fact that for any redshift the CHE channel has a small contribution to the intrinsic population of merging BBHs at $f_\mathrm{CHE}(z) < 0.2$. On The other hand the LIGO--Virgo detectable  $f^\mathrm{det}_{\chi_\mathrm{eff} > \chi_0}(z)$ manifests an almost flat behaviour up to $z = 0.6$ above which it sharply increases to reach unity at $z\simeq 1$. This sharp monotonic increase of $f^\mathrm{det}_{\chi_\mathrm{eff} > \chi_0}(z>0.6)$ is due to the contribution of tidally spun up BBHs formed from the CE channel completely dominates over BBHs formed from the SMT channel at $z>0.75$, as $f^\mathrm{det}_\mathrm{CE}(z>0.75) \gg f^\mathrm{det}_\mathrm{SMT}(z>0.75)$.

\begin{figure}
\centering
\includegraphics[width=0.5\textwidth]{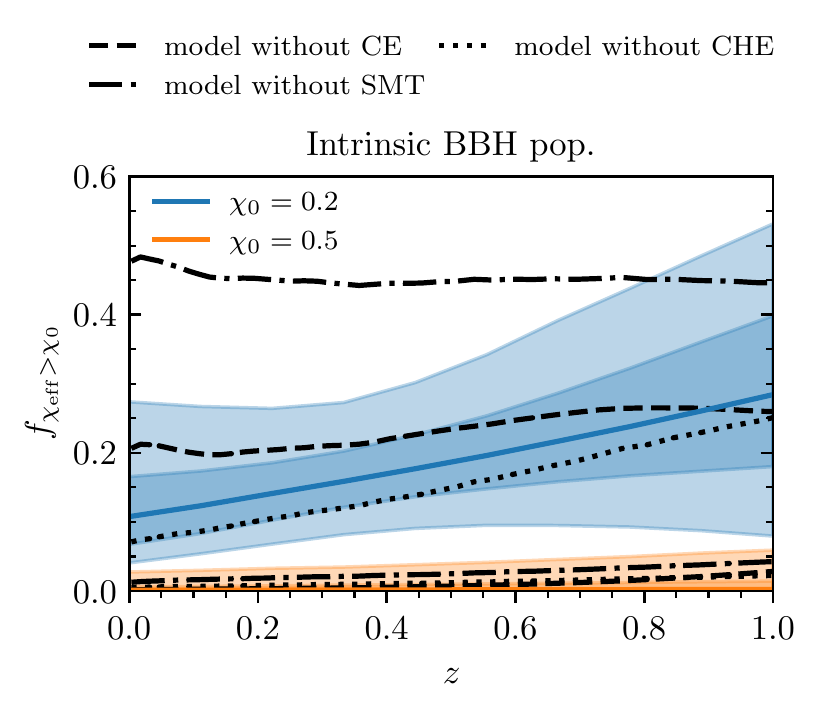}
\caption{Same as Figure~\ref{fig:underlying-fchieff-GWTC3} but we show the models excluding one of the three channels according to the legend.}
\label{fig:underlying-fchieff-GWTC3_without_channel}
\end{figure}

A comparison between the intrinsic $f_{\chi_\mathrm{eff} > 0.2}(z)$ when excluding one of the three field channels and the inferred distribution given the phenomenological model presented in Eq.~\eqref{eq:pop-chieff-z} is shown in Figure~\ref{fig:underlying-fchieff-GWTC3_without_channel}.
We can see that a model without the CHE channel is closer to the median inferred intrinsic fraction of $f_{\chi_\mathrm{eff}>0.2}$ than the fiducial model. Additionally, the model excluding the SMT channel is incompatible with the 90\% CI of the inferred fraction as it overpredicts the fraction of highly rotating BBHs.

%It is important to also notice that if our assumption of isolated binary evolution dominating the underlying BBH population is not correct, an additional channel with small $\chi_\mathrm{eff}\simeq 0$ BBHs would shift $f_{\chi_\mathrm{eff} > 0.2}(z)$ to lower values in agreement with the inferred evolution of Figure~\ref{fig:underlying-fchieff-GWTC3_without_CHE}. This last scenario might actually be the case if, e.g., dynamical formation in globular clusters with small BH birth spins contributes to say $\sim 10 \%$ to the total underlying BBH population \citep{2021ApJ...910..152Z}.

\end{document}